\documentclass[aoas]{imsart}

\RequirePackage{amsthm,amsmath,amsfonts,amssymb}
\RequirePackage[authoryear]{natbib}
\RequirePackage[colorlinks,citecolor=blue,urlcolor=blue]{hyperref}
\RequirePackage{graphicx}

\startlocaldefs
\theoremstyle{plain}

\theoremstyle{remark}

\newtheorem*{example}{Example}

\endlocaldefs

\usepackage{xcolor}
\usepackage[pagewise]{lineno}
\usepackage{rotating,multicol,multirow,array,styles}
\usepackage{float,tabularx,booktabs,longtable,subcaption,caption}
\usepackage[ruled,vlined,linesnumbered]{algorithm2e}

\newcommand{\barZ}{\bar{Z}}
\newcommand{\barz}{\bar{z}}

\usepackage{mathtools}

\let\tiny\normalsize
\let\scriptsize\normalsize
\let\large\normalsize
\let\Large\normalsize
\let\LARGE\normalsize
\let\huge\normalsize
\let\Huge\normalsize
\usepackage{enumitem}
\newcounter{dummy}
\makeatletter
\newcommand\myitem[1][]{\item[#1]\refstepcounter{dummy}\def\@currentlabel{#1}}
\makeatother

\usepackage[title,toc,titletoc,page]{appendix}

\setlist[enumerate]{wide=0pt, widest=99,leftmargin=\parindent, labelsep=*}

\newcolumntype{Y}{>{\centering\arraybackslash}X}
\usepackage{mathtools}
\DeclareFontFamily{U}{mathx}{\hyphenchar\font45}
\DeclareFontShape{U}{mathx}{m}{n}{<-> mathx10}{}
\DeclareSymbolFont{mathx}{U}{mathx}{m}{n}
\DeclareMathAccent{\widebar}{0}{mathx}{"73}
\makeatletter
\def\munderbar#1{\underline{\sbox\tw@{$#1$}\dp\tw@\z@\box\tw@}}
\makeatother

\newcommand\independent{\protect\mathpalette{\protect\independenT}{\perp}}
\def\independenT#1#2{\mathrel{\rlap{$#1#2$}\mkern2mu{#1#2}}}

\usepackage{secdot}

\makeatletter
\renewcommand\hyper@natlinkbreak[2]{#1}
\makeatother
\hypersetup{
	pdfborder = {0 0 0}
}

\allowdisplaybreaks

\usepackage{xr}
\makeatletter
\newcommand*{\addFileDependency}[1]{
  \typeout{(#1)}
  \@addtofilelist{#1}
  \IfFileExists{#1}{}{\typeout{No file #1.}}
}
\makeatother

\newcommand*{\myexternaldocument}[1]{
    \externaldocument{#1}
    \addFileDependency{#1.tex}
    \addFileDependency{#1.aux}
}

\myexternaldocument{./supplement}

\begin{document}

\begin{frontmatter}
\title{Evaluating time-varying treatment effects in hybrid SMART-MRT designs}
\runtitle{Hybrid SMART-MRT design}

\begin{aug}
\author[A]{\fnms{Mengbing}~\snm{Li}\ead[label=e1]{mengbing@umich.edu}\orcid{0000-0002-2264-8006}},
\author[B]{\fnms{Inbal Billie}~\snm{Nahum-Shani}\ead[label=e2]{inbal@umich.edu}\orcid{0000-0001-6138-9089}} 
\and
\author[A]{\fnms{Walter}~\snm{Dempsey}\ead[label=e3]{wdem@umich.edu}\orcid{0000-0002-6510-4121}}
\address[A]{Department of Biostatistics, University of Michigan\printead[presep={,\ }]{e1,e3}}
\address[B]{Institute for Social Research
University of Michigan\printead[presep={,\ }]{e2}}
\end{aug}

\begin{abstract}
Recently a new experimental approach -- the hybrid experimental design (HED) -- was introduced to enable investigators to  answer scientific questions about building behavioral interventions in which human-delivered and digital components are integrated and adapted on multiple timescales—slow (e.g., every few weeks) and fast (e.g., every few hours), respectively. An increasingly common HED involves the integration of the sequential, multiple assignment, randomized trial (SMART) with the micro-randomized trial (MRT), allowing investigators to answer scientific questions about potential synergistic effects of digital and human-delivered interventions.  Approaches to formalize these questions in terms of causal estimands and associated data analytic methods are limited. In this paper, we formally define and assess these synergistic effects in hybrid SMART-MRTs on both proximal and distal outcomes. Practical utility is shown through the analysis of M-Bridge, a hybrid SMART-MRT aimed at reducing binge drinking among first-year college students. 
\end{abstract}


\end{frontmatter}


\section{Introduction} \label{sec:introduction}

\label{sec:introduction:adaptive}
An adaptive intervention is an intervention approach that guides how dynamic information about the individual should be used in practice to make intervention decisions about the type, intensity, and modality of intervention delivery~\citep{collins2004conceptual, shani2012ExperimentalDesign}.
The goal is to address the unique and changing needs of individuals in a resource efficient manner~\citep{NahumShani2019Adaptive}.
Advances in digital technologies have enabled the rapid -- e.g., every minute~\citep{Battalio2021Sense2Stop} -- adaptation of interventions in real time to meet the immediate needs of individuals in daily life.
In behavioral health, mobile apps and wearable devices have presented new opportunities for adapting interventions to the individual’s rapidly changing state (e.g., emotions) and context (e.g., location) to improve positive behaviors (e.g., physical activity, mental health) or reduce negative ones (e.g., alcohol use, smoking)~\citep{klasnja2018EfficacyContextually,gustafson2014smartphone,ben2013development, riley2008internet}.
These just-in-time adaptive interventions (JITAIs;~\cite{nahumshani_annuvrev}) are delivered via automated, digital services (e.g., mobile devices), offering several advantages over human-delivered alternatives, including access, affordability, capacity to deliver complex intervention protocols with high fidelity, and the ability to address fast-changing conditions in everyday life~\citep{NahumShani2023DigitalAdaptive, Mohr2017PersonalSensing, NahumShani2018JITAI, Lattie2022AccessibleDigitalMH, Volkow2023SubstanceUseUpdate}. However, suboptimal engagement represents a major barrier to the effectiveness of digital services~\citep{Mohr2011SupportiveAccountability, Schueller2017EfficiencyModel, Yardley2016EngagementDigitalBCI}. Human-delivered services (e.g., by clinical staff) tend to be more engaging and produce larger effects~\citep{Mohr2011SupportiveAccountability, Schueller2017EfficiencyModel, Ritterband2009BehaviorChangeModel}. However, these services are adapted on a relatively slow timescale (e.g., every few weeks or months), are prone to inconsistent implementation, and are often more expensive and burdensome. Thus, integrating digital technologies with human-delivered support has enormous potential to increase the reach and impact of services for prevention and treatment in chronic illness populations.

Existing experimental designs and related data-analytic methods can be used to answer questions either about how to best employ components that are sequenced and adapted at relatively slow timescales (e.g., monthly) or about how to best employ components that are sequenced and adapted at much faster timescales (e.g., daily). However, these methodologies do not accommodate sequencing and adaptation of components at multiple timescales. Recently, the hybrid experimental design (HED) was introduced to close this gap.  HEDs provide a flexible framework to address this need by accommodating sequential randomization at both fast and slow timescales. Data from HEDs can be then be used by researchers to answer scientific questions about how to optimally blend digital and human-delivered intervention components~\citep{NahumShani2023DigitalAdaptive, nahum2022HybridExperimental}. In this paper, we focus on a particular type of hybrid design, the SMART-MRT hybrid design~\citep{NahumShani2024Hybrid}. The sequential multiple assignment randomized trial (SMART) implements sequential randomizations at slower timescales~\citep{Kidwell2023SMART}. Data from a SMART can be used to evaluate dynamic treatment regimens (DTRs;~\cite{Liu2014DTRSMART}), also called standard adaptive interventions in the behavioral literature~\citep{nahumshani_annuvrev}, in which intervention components are adapted on a relatively slow timescale. These DTRs define decision rules at each decision point, tailored to individuals’ time-varying characteristics and intermediate outcomes. The micro-randomized trial (MRT) involves frequent randomizations at fast timecales resulting in hundreds or thousands of decision points~\citep{klasnja2015microrandomized,dempsey2017StratifiedMicrorandomized}. Data from an MRT is used to assess effect moderation of digital interventions to inform JITAIs in which components are adapted in a fast timescale~\citep{nahum2022HybridExperimental}. The SMART-MRT hybrid design allows researchers to answer questions about how best to integrate digital intervention components that adapt rapidly with human-delivered components that adapt on a lower timescale. Current data analytic methods for SMART-MRT hybrid designs, however, focus only on separate analysis of the two components, treating the other component as a potential moderator of the others effectiveness but ignoring potential synergistic effects. 
Effect moderation analysis~\citep{boruvka2018assessing} is insufficient as it conditions on post-treatment variables (i.e., variables measured after baseline SMART randomization) and cannot be used to assess synergistic effects.
To fully realize the potential of integrating digital technologies and human-delivered support, formal definitions of synergistic effects and associated data analytic methods are critically needed.

\subsection{The M-Bridge Study and Existing Analyses} \label{sec:intro:mbridge}
The M-Bridge study employs a SMART-MRT hybrid design to reduce heavy drinking and related risks among first-year college students~\citep{patrick2020sequential}. The SMART involved two stages of randomization. First, students (N=591) were randomly assigned (with a 1:1 ratio) to one of two times for delivering an initial web-based intervention combining personalized normative feedback with bi-weekly self-monitoring of alcohol use: early (before the start of the fall semester), or later (during the first month of the fall semester). Second, participants who self-identified as heavy drinkers based on the bi-weekly self-monitoring (n=158; 26.7\%) were classified as non-responders and were re-randomized (1:1 ratio) to one of two strategies designed to bridge them to more intense treatment: either an email with available alcohol use intervention resources, or an invitation to interact with an online health coach. Self-monitoring ceased once a participant was identified as a non-responder. Those not identified as heavy drinkers (i.e., responders) continued with self-monitoring alone. The MRT involved bi-weekly randomization of those in the self-monitoring conditions to two types of prompts (1:1 ratio) encouraging participants to self-monitor their alcohol use: either a prompt emphasizing benefits to oneself (i.e., self-interest prompt) or a prompt emphasizing benefits to other (pro-social prompt). 

The study design of M-Bridge allows researchers to answer scientific questions about how to best blend three intervention components, two of which are delivered on a relatively slow timescale (i.e., the initial web-based intervention and subsequent bridging strategies) and one delivered on a faster timescale (i.e., bi-weekly self-monitoring prompts). Existing analyses of the M-Bridge study, however, are limited to analyzing either SMART~\citep{patrick2020sequential} or MRT~\citep{carpenter2023SelfrelevantAppeals} data in isolation. We refer to these causal effects as \emph{marginal effects} to emphasize that they marginalize over the other intervention component.  We refer to causal effects that jointly consider the two components as \emph{synergistic effects} to emphasize that they may look at contrasts in one component while keeping the other component to a fixed level.

In the statistical literature, various methods have been developed to analyze the causal effects of interventions in SMARTs and MRTs in separate contexts. For the analysis of data collected from a SMART, marginal mean models for estimating optimal dynamic treatment regimes (DTRs) have been established~\citep{murphy2001marginal, murphy_experimental_2005, orellana2010dynamic, chakraborty2014dynamic}. Building on this line of research, \cite{shani2012ExperimentalDesign} introduced the weighted and replicated (WR) approach for analyzing data from SMARTs where only a subset of individuals are re-randomized in the second stage of a SMART. Sample size calculation and power analysis for different types of SMARTs using the WR method have been introduced~\citep{seewald2020SampleSize}. For the analysis of data collected from an MRT, existing methods focus on estimating the time-varying causal excursion effect of binary treatments~\citep{dempsey2015randomised, liao2016sample}. The weighted and centered least squares (WCLS; \cite{boruvka2018assessing}) is regarded as the benchmark method used for estimating moderated causal excursion effects for a continuous outcome, with an extension to a binary outcome proposed in~\cite{qian2020EstimatingTimeVarying}. \cite{shi2023IncorporatingAuxiliary} and~\cite{10.1093/biomtc/ujaf129} improve asymptotic efficiency of WCLS by incorporating auxiliary variables and machine learning prediction algorithms respectively. 

The foundation for a data analytic method specific to the hybrid SMART-MRT was laid out by~\cite{NahumShani2023DigitalAdaptive}.  The current literature on data analytic methods for hybrid SMART-MRTs, however, has three improtant gaps:
(1) there is no formal statement of marginal and synergistic causal effects in a SMART-MRT within a causal framework;
(2) there is no robust statistical method that has both statistical consistency guarantees and ensures powerful test statistics for synergistic and marginal effects; and
(3) there is no comprehensive approach that simultaneously estimates the synergistic and marginal effects.

\subsection{Our Contributions} \label{sec:intro:contributions}
In this paper, we propose a novel data-analytic method for analyzing data from hybrid SMART-MRT studies that addresses these three gaps. 
Our four main contributions are summarized as follows. 
First, we formally define a set of causal estimands of scientific interest when analyzing data from a SMART-MRT hybrid design. 
These estimands include the interaction effects of human-delivered and digital components as well as main effects of one component averaging over the other. 
Second, a set of estimating equations is proposed that allow for simultaneous estimation of all causal estimands. Our method builds upon the WCLS method \citep{boruvka2018assessing} for MRTs and the WR method \citep{shani2012ExperimentalDesign} for SMARTs. 
We leverage similar ideas as in~\cite{10.1093/biomtc/ujaf129} to incorporate time-varying covariates to improve efficiency relative while avoiding potential causal bias when incorporating post treatment variables.  This leads to efficiency gains over WR methods for analyzing main DTR effects.
Third, we build a novel framework to incorporate eligibility when analyzing causal effects. 
Prior methods, such as WCLS, condition on individuals who are ``eligible'' to receive MRT treatment at a given decision point. Instead, we propose to average over eligibility status in estimating the treatment effects. 
Fourth, we apply our method to the M-Bridge study and draw scientific conclusions about the effects of relatively slow timescale (web-based interventions and subsequent bridging strategies) and fast timescale (self-monitoring prompts) intervention components in reducing binge-drinking.  Our analysis reveals potential synergistic effects and serves as a data analytic framework for future analysis of SMART-MRT hybrid studies.

The rest of the paper is organized as follows. Section \ref{sec:method} describes a typical hybrid SMART-MRT design and introduces notation. 
Section \ref{sec:estimation} lays out the modeling assumption and inference framework. 
Section \ref{sec:sim} compares estimation performances of the proposed and alternative approaches via simulation studies. 
Section \ref{sec:data} applies the proposed method to the M-Bridge study.
Section \ref{sec:discussion} concludes with a brief discussion on study limitations and future directions.

\section{Preliminaries} \label{sec:method}
Motivated by the M-Bridge study, we start by introducing our general notation for a two-stage hybrid SMART-MRT.  Adaptation to similar hybrid SMART-MRT designs is possible and will be discussed below.  Specifically, we will discuss additional considerations for application of our proposed approach to the M-Bridge study in Section \ref{sec:data}. Table~\ref{tab:notation} in the Supplementary Materials summarizes all notation for defining the data, estimands, and estimators related to a SMART-MRT hybrid design.

\subsection{Study Design and Notation} \label{sec:method:notation}

\begin{figure}[ht]
    \centering
    \includegraphics[width=1.0\linewidth]{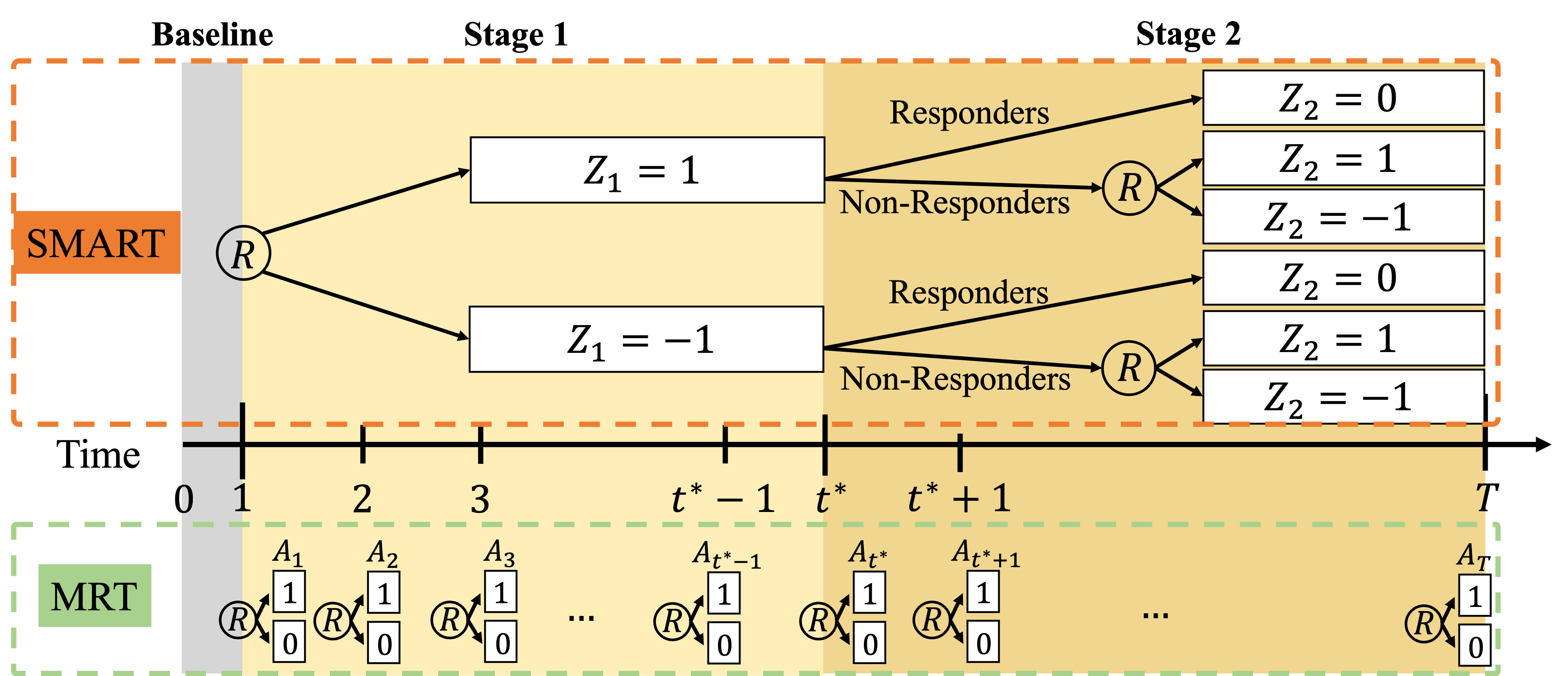}
    \caption{A two-stage SMART-MRT hybrid design over $T$ time points. Baseline information is collected time 0. The circled Rs represents randomization events.
    Upper panel: the restricted SMART component of the hybrid design where only non-responders are re-randomized in Stage 2. Lower panel: the MRT component of the hybrid design that randomizes a binary intervention at each time point.}
    \label{fig:hybrid_design_toy}
\end{figure}

\begin{figure}[ht]
    \centering
    \includegraphics[width=0.9\linewidth]{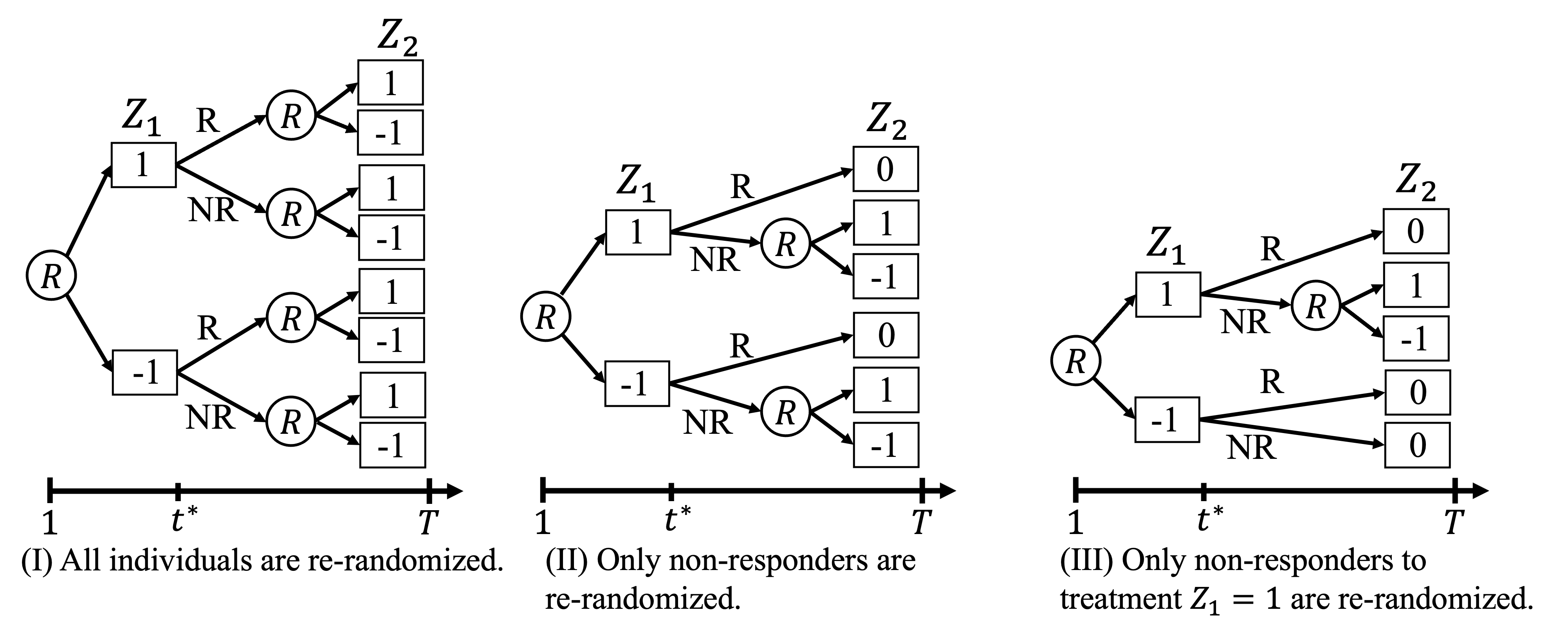}
    \vspace{-1em}
    \caption{Three commonly used two-stage SMART designs in \cite{seewald2020SampleSize}. Non-circled Rs and NRs are short for responders and non-responders. Circled Rs represent randomization. }
    \label{fig:smart:designs}
\end{figure}

Let $X_0$ denote a vector of baseline covariates. 
At the beginning of Stage 1 of the study, an individual is randomly assigned to the first-stage intervention $Z_1 \in \cZ_1 := \{-1, 1\}$. 
During Stage 1, the individual is subsequently randomized to an intervention $A_t \in \{0, 1\}$ at each time point $t = 1, \ldots, t^*-1$.
Transition to Stage 2 occurs at time $t = t^*$, at which time individuals are randomized to a second-stage intervention $Z_2 \in \cZ_2$. Here $\cZ_2$ can be $\{-1, 1\}$ or $\{-1, 0, 1\}$ depending on whether we consider an unrestricted or a restricted SMART design, i.e., whether every individual is re-randomized or not (see Figure \ref{fig:smart:designs} for three common SMART designs).
During Stage 2, the individual is randomized to intervention $A_t \in \{0, 1\}$ at each time point $t = t^*+1, \ldots, T$.
Individual and contextual information at the $t$-th time point is represented by $X_t$ and is measured before receiving intervention $A_t$.
The proximal response, denoted $Y_{t+1}$, is observed after receiving intervention $A_t$.  
Additionally, let $Y$ (with no subscript) be the distal outcome at the end of the study.
The proximal response measures near-term impact of intervention components, while the distal outcome is designed to measure longer-term impact of the sequence of interventions.
The collection of observed data is 
$$
\underbrace{X_0}_{\text{Baseline}}, 
\underbrace{Z_1, X_1, A_{1}, Y_{2}, \ldots, X_{t^*-1}, A_{t^*-1}, Y_{t^*-1}}_{\text{Stage 1}}, 
\underbrace{Z_2, X_{t^*+1}, A_{t^*+1}, Y_{t^*+1}, \ldots, X_T, A_{T}, Y_{T+1}}_{\text{Stage 2}}.$$
Figure \ref{fig:hybrid_design_toy} presents a restricted two-stage hybrid SMART-MRT design. Table~\ref{tab:notation} provides a summary of the above notation.
For brevity, we refer to $(Z_1,Z_2)$ as the \emph{short-time scale (STS)} intervention component and $(A_1,\ldots, A_T)$ as the \emph{fast-time scale (FTS)} intervention component.

An overbar denotes a sequence of random variables (uppercase letters) or realized values (lowercase letters) through a specific intervention occasion. For example, $\widebar{A}_t = (A_1, \ldots, A_t)$, $\widebar{X}_t = (X_0, X_1, \ldots, X_t)$, and $\widebar{Y}_{t+1} = (Y_2, \ldots, Y_{t+1})$. 
For notational convenience, $\widebar Z_t$ denotes the sequence of FTS interventions in stages prior to time $t$, i.e., $\widebar{Z}_t = Z_1$ if $t < t^*$ and $\widebar{Z}_t = (Z_1, Z_2)$ if $t \geq t^*$. 
The complete history of observable information up to $t$ is $H_t = \left( \widebar{X}_t, \widebar{Z}_t, \widebar{A}_{t}, \widebar{Y}_t \right)$.

Next, we introduce the randomization scheme used in a SMART-MRT. In the embedded SMART, the Stage 1 randomization probability is $P(Z_1 = z_1 \mid X_0)$ for $z_1 \in \cZ_1$, i.e., the randomization depends only on baseline information. In Stage 2, the randomization probability is $P(Z_2 = z_2 \mid R)$ for $z_2 \in \cZ_2$, i.e., only depends on binary response status $R$ which is a deterministic function of the observed history up to time $t^* -1$, $H_{t^*-1}$. For the embedded MRT component, the FTS intervention randomization probability is $P (A_t = a \mid H_t) = p_t (a \mid H_t)$ for $a \in \{0,1\}$, i.e., depends on the observed history which includes prior STS interventions.

In the M-Bridge study, recall that first-stage STS interventions are early ($Z_1 = 1)$ and later ($Z_1 = -1)$ with a 1:1 ratio, i.e., $P(Z_1 = 1 | X_0)=1/2$. 
FTS interventions are digital SI prompts ($A_t = 1$) and PS prompts ($A_t = 0$) delivered prior to biweekly self-monitoring surveys with probability $p_t (1 \mid H_t) = 0.5$. 
The binary response indicator $R$ is whether a student is classified as a non-heavy drinker based on self-monitoring surveys: heavy drinkers ($R = 0$) are considered \textit{non-responders} to $Z_1$.
For heavy drinkers, second-stage STS interventions are a resource email ($Z_2 = 1$) or an online health coach ($Z_2 = -1$) also with equal probability. In other words, $P(Z_2 = z_2 \mid R = 0) = 1/2$ for $z_2 = -1, 1$ and $P(Z_2 = 0 \mid R = 1) = 1$.
The proximal outcome $Y_{t+1}$ is the maximum number of alcoholic drinks consumed within 24 hours during the past two weeks of the $t$-th SM survey. 
The distal outcome $Y$ is the cumulative number of alcoholic drinks consumed throughout the study.

\begin{remark} \label{remark:smartdesign}
The embedded SMART in the M-Bridge study is a restricted SMART design -- type (II) in Figure~\ref{fig:smart:designs} below. Other commonly used SMART designs are illustrated in Figure \ref{fig:smart:designs}; see \cite{patrick2020sequential,shani2023DesignExperiments} for more examples. The method proposed in this paper applies to all these SMART designs in a hybrid SMART-MRT study, although we focus on type (II) based on the motivating M-Bridge study. 
\end{remark}


\subsection{Potential outcomes and dynamic treatment regimes}
To define causal estimands, we adopt the potential outcome framework \citep{rubin1974EstimatingCausal,robins2000SensitivityAnalysis}. 
Denote $Y_{t+1} \left( \widebar z_t, \widebar a_{t} \right)$ as the potential outcome for the proximal response under a specific STS intervention sequence $\widebar z_t$ and FTS intervention sequence $\widebar a_{t}$ up to time $t$. Similarly, let $A_{t} \left( \widebar z_t, \widebar a_{t-1} \right)$, $X_{t} \left( \widebar z_t, \widebar a_{t-1} \right)$, and $H_t \left(\widebar z_t, \widebar a_{t-1} \right)$ be the potential outcomes for the FTS intervention, covariates, and history, respectively.


c
A dynamic treatment regime (DTR) is a sequence of decision rules $\widebar d = (d_1, d_{2})$ embedded in the SMART. The decision rule $d_1$ is a mapping from $X_0$ to the first-stage STS intervention space $\cZ_1$, and $d_2$ is a mapping from $H_{t^*-1}$ to $\cZ_2$.
Let $\cD = \{\bar d = (d_1, d_2): d_k \in \cZ_k, k = 1,2\}$ be the collection of all possible DTRs. 
The potential proximal outcome under a regime $\widebar d$ and FTS intervention sequence $\widebar a_t$ is defined as 
\begin{align}
& Y_{t+1} \left( \widebar d, \widebar a_{t} \right) =
\sum_{z_1 \in \cZ_1} 1\{z_1 = d_1 (X_0)\} \sum_{z_2 \in \cZ_2} 1\{z_2 = d_2 (H_{t^*-1})\} 
Y_{t+1} \left( \widebar z, \widebar a_{t} \right). 
\end{align}
While our proposed methodology can be applied broadly to estimands under any DTR, in line with secondary analyses of SMARTs \citep{seewald2020SampleSize} we will focus on DTRs where $d_1$ does not depend on~$X_0$, and~$d_2$ only depends on response status. In the M-Bridge study, the restricted SMART design contains four DTRs that $\cD = \{-1, 1\}^{\times 2}$ and an individual has equal probability $P(\widebar d = (l_1, l_2)) = 1/2 \times 1/2 = 1/4$ for $\forall (l_1, l_2) \in \cD$ to be consistent with any of the DTRs.  

\subsection{Causal Estimands} \label{sec:method:effects}

In this section, we define causal estimands and build associated estimators focusing on the sequence of proximal outcomes. Similar estimands and estimators can be defined and built for distal outcomes; however, these require nuanced considerations around delayed effects as was discussed in~\cite{qian2025distal}.  Focusing on proximal outcomes helps to clarify how we define and estimate synergistic effects which can then be translated in future work to similar synergisitic effects for distal outcomes.


We motivate our causal estimands from four scientific questions about the effects of the slow-time scale (STS) and fast-time scale (FTS) intervention components on proximal outcomes (bi-weekly reported maximum drinks) in the M-Bridge study:
(1) Fixing the user to a prompt emphasizing benefits to oneself (self-interest prompt), is it better to initiate the web-based intervention early and use a resource email to bridge non-responders, or initiate the web-based intervention later and use an online coach?
(2) Fixing the user to receive an early intervention initiation and an email-based bridging strategy, is it more effective to deliver a self-interest prompt or a pro-social prompt?
(3) Averaging over prompt type (self-interest and pro-social), is it more effective to initiate the intervention early and use a resource email to bridge non-responders, or initiate the intervention later and use a resource email?
(4) Averaging across all STS intervention sequences, is a self-interest prompt more effective than a pro-social prompt in a given week?
Questions (1) and (2) correspond to the synergistic effects between the STS and FTS intervention components, which capture how the impact of one intervention component (e.g., timing of the web-based intervention) interacts with the other intervention component (e.g., the type of prompt delivered).
Questions (3) and (4) correspond to the main effects of one intervention, which is the difference in the mean outcome between different levels of an intervention, averaged across all other intervention components \citep{collins2009design}. Additional scientific questions about the proximal interaction effects in this hybrid design are listed in Table \ref{tab:questions} of Appendix \ref{sec:questions}. Synergistic effects (Questions 1 and 2) are distinct from moderation effects~\citep{boruvka2018assessing,dempsey2017StratifiedMicrorandomized} as moderation analyses conditions on previous interventions while our questions consider fixed regimes.  Question 2, for example, is the effect of self-interest prompt versus pro-social prompt under a fixed dynamic treatment regime (DTR), which is distinct from an effect that is conditional on the STS interventions delivered to an individual.

We define the causal estimands motivated by these questions as follows:
\begin{enumerate}
\myitem[(I.D)] \label{question:interaction:Z} \hspace{0.5em} (\textbf{I}nteraction effect for \textbf{D}TRs) The proximal effect comparing two DTRs $\widebar d$ versus $\widebar d'$ when assigning a fixed FTS intervention $A_t = a$ at time point $t$:
\begin{equation} \label{eq:question:interaction:Z}
\EE \left[ Y_{t+1} \left( \widebar d, (\widebar{A}_{t-1}, a) \right) - Y_{t+1} \left( \widebar d', (\widebar{A}_{t-1}, a) \right) \mid X_0 \right].
\end{equation}

\myitem[(I.A)] \label{question:interaction:A} \hspace{0.5em} (\textbf{I}nteraction effect for $\bA_t$) The marginal proximal effect of FTS interventions at time point $t$, i.e., $A_t = 1$ versus $A_t = 0$, when assigning a fixed DTR $\widebar d$:
\begin{equation} \label{eq:question:interaction:A}
\EE \left[ Y_{t+1} \left( \widebar d, (\widebar{A}_{t-1}, 1) \right) - Y_{t+1} \left( \widebar d, (\widebar{A}_{t-1}, 0) \right) \mid X_0 \right].
\end{equation}

\myitem[(A.D)] \label{question:average:D} \hspace{0.5em} (\textbf{A}veraged effect for \textbf{D}TRs) The marginal proximal effect of two DTRs $\widebar d$ versus $\widebar d'$ at time point $t$, averaging over FTS interventions:
\begin{equation} \label{eq:question:average:D}
\EE \left[ Y_{t+1} \left( \widebar d, \widebar{A}_{t} \right) - Y_{t+1} \left( \widebar d', \widebar{A}_{t} \right) \middle\vert X_0 \right].
\end{equation}

\myitem[(A.A)] \label{question:average:A} \hspace{0.5em} (\textbf{A}veraged effect for $\bA_t$) The marginal proximal effect of FTS interventions  at time point $t$, i.e., $A_t = 1$ versus $A_t = 0$, averaging over all DTRs:
\begin{equation} \label{eq:question:average:A}
\sum\nolimits_{(l_1, l_2) \in \cD} P(\widebar d = (l_1, l_2)) \EE \left[ Y_{t+1} \left( (l_1, l_2), (\widebar{A}_{t-1}, 1) \right) - Y_{t+1} \left( (l_1, l_2), (\widebar{A}_{t-1}, 0) \right) \middle\vert X_0 \right].
\end{equation}
\end{enumerate}

The expectations are taken with respect to the distribution of the potential history $H_t (\widebar d, \widebar A_{t-1})$ given $X_0$, under the DTRs and MRT randomization probability.
The causal estimands average over past FTS interventions $\widebar A_{t-1}$ and time-varying variables in the history except for a subset of baseline variables $X_0$.
This averaging mitigates the large space of FTS intervention sequences $\widebar{a}_t \in \{0, 1\}^T$ due to large $T$ relative to the number of observations. This approach is taken from the existing MRT literature \citep{boruvka2018assessing,shi2023IncorporatingAuxiliary,qian2020EstimatingTimeVarying} which also averages over prior interventions when defining ``causal excursion effects''.
While it is possible to define causal effects for specific intervention sequences of short length, we focus on average effects to maintain interpretability and avoid the complexity that comes with modeling and estimating effects over such intervention sequences. Extensions to handle such complexity are possible; see~\cite{10.1093/biomtc/ujaf129} for how to estimate effects of intervention sequences on proximal outcomes.

We next express the proximal effects in terms of the observed data, by assuming positivity, consistency, and sequential ignorability \citep{robins1994CorrectingNoncompliance,robins1997CausalInference}:
\begin{assumption} (Causal Identification) \label{assumption:identification}
\textcolor{white}{   }
\begin{itemize}
\item Positivity: $P(Z_1 = z_1 \mid X_0) > 0$, $P(Z_2 = z_2 \mid R, Z_1 ) > 0$, and $P(A_t = a \mid H_t) > 0$ almost everywhere for all $z_1 \in \cZ_1, z_2 \in \cZ_2, a \in \cA$.

\item Consistency: for each $t \leq T$, $\left\{ X_{t} (\widebar{Z}_t, \widebar A_{t-1}), A_{t} (\widebar{Z}_t, \widebar A_{t-1}), Y_{t+1} (\widebar{Z}_t, \widebar A_{t}) \right\} = \{ X_{t}, A_{t}, Y_{t+1} \}$. 

\item Sequential ignorability (SI): 
\begin{enumerate}[label=(\alph*)]
    \item \hspace{0.5em} $\left\{ X_{t+1} (\widebar{z}_t, \widebar a_{t}), A_{t+1} (\widebar{z}_t, \widebar a_{t}), Y_{t+1} (\widebar{z}_t, \widebar a_{t}), \right. \allowbreak \left. 
    \ldots, Y_{T+1} (\widebar{z}_T, \widebar a_{T}) \right\} \independent \widebar Z_t \mid H_t \setminus \{\widebar Z_t\}$

    \item \hspace{0.5em} $\{Y_{t+1} (\widebar{z}_t, \widebar a_{t}), X_{t+1} (\widebar{z}_t, \widebar a_{t}), A_{t+1} (\widebar{z}_t, \widebar a_{t}), \ldots, Y_{T+1} (\widebar{z}_t, \widebar a_{T}) \} \independent A_t \mid H_{t}$. 
    
\end{enumerate}
\end{itemize}
\end{assumption}

The positivity assumption implies that an individual has a positive probability to follow any DTR and receive any FTS intervention given the history. In a hybrid SMART-MRT study, the SMART and MRT randomization probabilities are known at all decision points $t = 1, \ldots, T$.
The consistency assumption subsumes Rubin's \textit{Stable Unit Treatment Value Assumption} (SUTVA) that no interference exists between individuals~\citep{rubin1980randomization}.
SI(a) and (b) are commonly seen in standard SMART and MRT studies. 
By study design, SI is automatically satisfied.
Under Assumption \ref{assumption:identification}, we have
\begin{align}
&\ \EE \left[ Y_{t+1} \left( \widebar d_t, (\widebar{A}_{t-1}, a) \right) \middle\vert X_0 \right] \nonumber \\
=& \ \mathbb{E} \left[ \mathbb{E} \left[ \mathbb{E} \left[ 
\mathbb{E} \left[ Y \mid H_t, A_{t} = a \right] \middle\vert H_{t^*-1}, Z_{2} = d(H_{t^*-1}) \right] \middle\vert X_0, Z_{1} = d_1(X_0) \right] 
\middle\vert X_0 \right] \label{eq:marginal:mean:expression:conditional} \\
=& \ \mathbb{E} \left[ 
\sum_{z_1 \in \cZ_1} \sum_{z_2 \in \cZ_2} 
\frac{ 1 \{z_1 = d_1 (X_0) \} 1 \{z_2 = d_2 (H_{t^*-1}) \}  }{
P(Z_1 = z_1, Z_2 = z_2 \mid H_{t^*-1})}
\frac{ 1 \{A_t = a \}}{p_t(a \mid H_t,\widebar Z = \widebar z)}
Y \middle\vert X_0 \right]. \label{eq:marginal:mean:expression:weighted}
\end{align}
The proof of \eqref{eq:marginal:mean:expression:conditional} and \eqref{eq:marginal:mean:expression:weighted} can be found in Appendix~\ref{sec:proof:observeddata}.

\section{Estimation and Inference} \label{sec:estimation}

\subsection{Modeling Assumptions} \label{sec:estimation:model}

In the following, we propose a method to jointly estimate marginal interaction and main effects of the DTRs and FTS interventions. 
Based on \eqref{eq:marginal:mean:expression:conditional}, we assume that the expectation of the proximal outcome given STS and FTS intervention assignments takes the form
\begin{align}
\EE \left[ Y_{t+1} \left( \widebar d, (\widebar{A}_{t-1}, a) \right) \right] &= (a - \rho) f_t (\widebar d)^\top \beta + m_t(\widebar d)^\top \eta, \label{eq:marginal:model:eq1}
\end{align}
where $f_t (\widebar d)$ and $m_t(\widebar d)$ are a $p$- and $q$-dimensional vector functions of DTRs $\widebar d$, respectively. Here, $\rho \in (0, 1)$ is a fixed pseudo-centering probability for FTS interventions. As will be discussed in Section \ref{sec:estimation}, $\rho$ should be chosen to stabilize the estimation of $(\beta, \eta)$. 
In addition, by definition of $\widebar d$, $f_t$ and $m_t$ should depend on only $d_1$ if $t < t^*$, but may depend on $d_1$ and $d_2$ if $t \geq t^*$. 
Moreover, while the regression coefficients $\beta$ and $\eta$ are constant over time, time-varying effects may be included through time-dependent components (e.g., a linear term in time $d_1 t$) into $f_t$ and $m_t$. 
If we are interested in the conditional expectation given baseline variables $X_0$, i.e., $\EE \left[ Y_{t+1} \left( \widebar d, (\widebar{A}_{t-1}, a) \right) \middle\vert X_0 \right]$, we may incorporate $X_0$ into $f_t$ and $m_t$.

The functions $f_t$ and $m_t$ may differ depending on whether the marginal effect of DTRs, averaged over $\widebar A_t$, is expected to remain constant across stages. As will be illustrated in Example \ref{ex:marginal:model}, we set $f_t = m_t$ when both MRT and SMART randomization probabilities are constant (see Equation \eqref{eq:example:model:interaction}). 
In contrast, Simulation Scenario II in Section \ref{sec:sim} considers MRT randomization probabilities that depend on prior STS interventions, leading to stage-specific marginal effects of $d_1$ averaged over $\widebar A_t$. To accommodate this, $m_t$ includes stage-specific intercepts and $d_1$ coefficients, while $f_t$ shares these terms across stages. Although stage-specific terms could also be included in $f_t$ as well in Example \ref{ex:marginal:model}, doing so would reduce efficiency of estimating the proximal intervention effects, since stage-specific terms would be estimated using only data from the corresponding stage. We thus maintain separate specifications for $f_t$ and $m_t$ to reflect stage-wise differences of the contributions of DTRs to the proximal effects while preserving estimation efficiency.

Using \eqref{eq:marginal:model:eq1}, the interaction effects and the average effects of FTS interventions discussed in Section \ref{sec:method:effects} can be expressed using the coefficients $(\beta, \eta)$ as the followings:
\begin{align}
\text{ \ref{question:interaction:Z}} \ \ \ & \left[(1 - \rho) f_t (\widebar d) - \rho f_t (\widebar d') \right]^\top \beta + \left( m_t(\widebar d) -m_t(\widebar d') \right)^\top \eta \\
\text{ \ref{question:interaction:A}} \ \ \ & f_t (\widebar d)^\top \beta \\
\text{ \ref{question:average:A}} \ \ \ & 
\sum_{(l_1, l_2) \in \cD} 
P(\widebar d = (l_1, l_2))
f_t ((l_1, l_2))^\top \beta.
\end{align}

On the other hand, the average effects of DTRs marginalized over FTS interventions, in \ref{question:average:D}, may not be directly attainable from \eqref{eq:marginal:model:eq1} if the MRT randomization probability $p_t(a \mid H_t)$ depends on time-varying moderators in the history. 
In fact, $\eta$ is interpreted as the proximal effects comparing DTRs averaging over FTS interventions, \textit{as if} all FTS interventions were randomized with probability $p_t(1 \mid H_t) = \rho$ (see Example \ref{ex:marginal:model}).
Therefore, we obtain this effect directly by projecting onto the space of $m_t(\widebar d)$ as
\begin{align}    
\EE \left[ Y_{t+1} \left( \widebar d, (\widebar{A}_{t-1}, A_t) \right) \right] &= m_t(\widebar d)^\top \gamma, \label{eq:marginal:model:eq2}
\end{align} 
where $\gamma \in \RR^q$. 
Here the coefficient $\gamma$ is interpreted as the proximal effect comparing DTRs averaging over FTS interventions, under the actual MRT randomization probability by study design. 
See Section \ref{sec:marginal:model:eq2} of the Appendix for more detailed discussion.

We now give a simple example to illustrate the model and interpretation of the coefficients. 

\begin{example} \label{ex:marginal:model}
Consider the two-stage SMART-MRT hybrid design shown in Figure \ref{fig:hybrid_design_toy}. 
All individuals are randomized with equal probabilities to one of the two Stage-1 STS interventions, and only non-responders are randomized with equal probabilities to Stage-2 STS interventions.
As a result, each individual is assigned to one of the four DTRs with equal probability.
To illustrate key estimands of interest in hybrid designs, we consider a simplified working model for the marginal expectation of the proximal outcome:
\begin{align}
\begin{split} \label{eq:example:model:interaction}
\EE \left[ Y_{t+1} \left( \widebar d = (l_1, l_2), (\widebar{A}_{t-1}, a_t) \right) \right] & =
(a_t - 1/2) \left( \beta_0 + \beta_1 l_1 + \beta_2 \delta_t l_2 + \beta_3 \delta_t l_1 l_2 \right) \\
&\ \ \ + \eta_0 + \eta_1 l_1 + \eta_2 \delta_t l_2 + \eta_3 \delta_t l_1 l_2 \\
E[ Y_{t+1} ((d_1, d_2) = (l_1, l_2), \bar A_{t}))] &= \gamma_0 + \gamma_1 l_1 + \gamma_2 \delta_t l_2 + \gamma_3 \delta_t l_1 l_2. 
\end{split}
\end{align}
where $\delta_t = I\{t > t^*\}$ is a Stage 2 indicator. 
These models are simplified and time-invariant for illustrative purposes. In practice, time-varying components can be accommodated (e.g., replace $\beta_1 d_1$ with $\beta_1 d_1 + \beta_1' d_1 t$). 
Alternatively, even if \eqref{eq:example:model:interaction} is misspecified for true time-varying effects, we may still view \eqref{eq:example:model:interaction} as a working model targeting summaries of the time-varying effects.
For \ref{question:interaction:Z}, the effect of comparing $\bar d = (1, 1)$ versus $\bar d' = (-1, 1)$ when $A_t = 1$ at $t > t^*$ is given by $\beta_1 + \beta_3 + 2 \eta_1 + 2 \eta_3$. 
For \ref{question:interaction:A}, the effect of $A_t = 1$ versus $A_t = 0$ when $\bar d = (1, 1)$ at $t \geq t^*$ is given by $\beta_1 + \beta_3$. 
For \ref{question:average:A}, the effect of $A_t = 1$ versus $A_t = 0$ averaging over DTRs at $t > t^*$ is given by 
\begin{align*}
\sum_{l_1, l_2 \in \{-1, 1\}^2} & P(\widebar d = (l_1, l_2)) \left[ \left( 1 - \frac{1}{2} \right) \left( \beta_0 + \beta_1 l_1 + \beta_2 l_2 + \beta_3 l_1 l_2 \right) + \eta_0 + \eta_1 l_1 + \eta_2 l_2 + \eta_3 l_1 l_2 \right.\\
&\ \ \ - \left. \left( 0 - \frac{1}{2} \right) \left( \beta_0 + \beta_1 l_1 + \beta_2 l_2 + \beta_3 l_1 l_2 \right) - (\eta_0 + \eta_1 l_1 + \eta_2 l_2 + \eta_3 l_1 l_2) \right] = \beta_0.
\end{align*}
The effect of DTRs $\bar d = (1, 1)$ versus $\bar d' = (-1, 1)$ averaging over all past FTS interventions is then
$2\gamma_1 + 2\gamma_3$. 
Table \ref{tab:questions} in Appendix \ref{sec:questions} expresses additional marginal effects of interest.
\end{example}

\subsection{Estimation and Inference} \label{sec:estimand:estimation}
We now describe our two-step approach for estimation of the parameters $\theta := (\beta, \eta, \gamma)$. The estimation procedure is designed to leverage the hybrid SMART-MRT structure by incorporating both time-varying and stage-specific intervention assignments.
Broadly, Step 1 estimates the proximal intervention effects $(\beta, \eta)$ using a weighted estimating equation inspired by WCLS, while we improve efficiency by incorporating auxiliary variables that moderate intervention effects.
Step 2 estimates the effects of DTRs embedded in $\gamma$, by regressing predicted outcomes from Step 1 on the regime indicators, using a method inspired by the WR approach.
Overall, we appropriately propagate the uncertainty in the two steps by deriving the asymptotic distribution for $\theta$. 

\paragraph*{Step 1: Estimating $(\beta, \eta)$}
We begin by modeling the marginal expectation of the proximal outcome $Y_{t+1}$ as in \eqref{eq:marginal:model:eq1}. Let $g_t(H_t)$ denote an $r$-dimensional vector function of the history $H_t$ that will be used as control variables. 
Let $S_t$ denote a set of auxiliary variables (dimension $l \leq r$) that are believed to be effect moderators. We assume $S_t \subseteq g_t(H_t)$ and $S_t \cap f_t(\bar d) = \emptyset$ to ensure the auxiliary variables serve as valid augmentation terms.
To appropriately handle the hybrid design, we assign each observation two types of weights.
The first is a SMART weight associated with a DTR $\bar d$ as 
\begin{align}
W_{\widebar d}^S = 
\frac{ 1 \{Z_1 = d_1 (X_1) \} 1 \{Z_2 = d_2 (H_{t^*-1}) \} }{P(Z_1 \mid X_0) P(Z_2 \mid R)},
\end{align}
which accounts for the individual's consistency with regime $\bar d = (d_1, d_2)$ under the SMART component of the design.
The second is an MRT weight associated with $A_t$ defined as
\begin{align}
W_t^M = \frac{\tilde{p}_t(A_t)}{p_t (A_t \mid H_t)}, 
\end{align}
where the numerator $\tilde{p}_t(A_t) = \rho^{A_t} (1-\rho)^{1 - A_t}$ and $\rho \in (0, 1)$ is the constant in \eqref{eq:marginal:model:eq1}. This MRT weight adjusts for the randomization probability in the MRT component at time $t$.

We then define the following weighted and centered estimating equation
\begin{align}
\begin{split} \label{eq:ee:beta}
& U_1 (\alpha_0, \alpha_1, \beta, \eta) = 
\sum_{\bar d \in \cD} \sum\limits_{t=1}^T 
W_{\widebar d}^S W_t^M
\left[ Y_{t+1} - \left( g_t(H_t) - \mu_{t,\widebar d} (H_t) \right)^\top \alpha_0 \right.  \\
&\ \ - \left.(A_t - \rho) \left( f_t (\widebar{d})^\top \beta + (S_t - \psi_t(\widebar{d}))^\top \alpha_1 \right) 
- m_t (\widebar{d}, s)^\top \eta 
\vphantom{\left( g_t(H_t) - \mu_{t,\widebar d} (H_t) \right)^\top} \right] 
\begin{pmatrix}
g_t(H_t) - \mu_{t,\widebar d} (H_t) \\
(A_t - \rho) f_t \left( \widebar{d} \right) \\
(A_t - \rho) (S_t - \psi_t(\widebar{d})) \\
m_t \left( \widebar{d} \right)
\end{pmatrix}.
\end{split}
\end{align}
Here, $\mu_{t,\bar d}(H_t)$ and $\psi_t(\bar d)$ are centering functions designed to make estimating equations for the intervention effect parameters $(\beta, \eta)$ orthogonal to the nuisance parameters $(\alpha_0, \alpha_1)$.  
Specifically, the two centering functions satisfy:
\begin{align*}
\EE \left[ W_{\bar d}^S W_t^M (g_t(H_t) - \mu_{t, \bar d}(H_t)) \mid \bar Z \right] = 0, \\
\
\EE \left[ W_{\bar d}^S W_t^M (A_t - \rho)^2 (S_t - \psi_t(\bar d)) \mid \bar Z \right] = 0.
\end{align*}
While various choices for the centering functions are available, we assume a convenient choice in the rest of this paper as
\begin{equation} \label{eq:center}
    \mu_{t,\widebar d} (H_t) = \frac{\sum_{t=1}^T W_{\widebar d}^S g_t(H_t)}{\sum_{t=1}^T W_{\widebar d}^S }, \ 
    \psi_t(\widebar{d}, s) \equiv \psi = \frac{\sum_{\bar d \in \cD} \sum_{t=1}^T 
W_{\widebar d}^S
\tilde p_t(1) (1 - \tilde p_t(1)) S_t}
{\sum_{\bar d \in \cD} \sum_{t=1}^T W_{\widebar d}^S \tilde p_t(1) (1 - \tilde p_t(1))}. 
\end{equation}
The estimates $(\hat \alpha_0, \hat \alpha_1, \hat \beta, \hat \eta)$ are then obtained by solving $\mathbb{P}_n U_1(\alpha_0, \alpha_1, \beta, \eta) = 0$.

\begin{remark}[Centering] \label{remark:ee:center}
The control variables in \eqref{eq:ee:beta} are chosen such that $g_t(H_t)$ is a working model for $\EE \left[ W_{\widebar d}^S W_t^M Y_{t+1} \mid H_t \right]$. 
Unlike the WCLS method, which uses $g_t(H_t)^\top \alpha$ without centering, our approach centers $g_t(H_t)$ around its conditional mean given the DTRs. This centering is essential for unbiased estimation of the interaction effects comparing DTRs at a fixed $A_t = a$ (see \ref{question:interaction:Z}), even if the nuisance model for $\EE \left[ W_{\widebar d}^S W_t^M Y_{t+1} \mid H_t \right]$ is misspecified, as shown in Section \ref{sec:sim}.
Similarly, we impose an orthogonality condition on the auxiliary variables to ensure unbiased but more efficient estimation of $(\beta, \gamma)$ under a time-varying moderator $S_t$. The idea of incorporating auxiliary variables has been discussed by \cite{shi2023IncorporatingAuxiliary} to account for time-varying intervention effects in MRT studies.
Orthogonality ensures consistent causal estimation while allowing us to incorporate control variables~$g_t(H_t)$ and auxiliary variables~$S_t$ that can improve statistical efficiency. 
\end{remark}

\begin{remark}[Weights for Hybrid Designs]
The SMART weight $W_{\widebar d}^S$ in \eqref{eq:ee:beta} accounts for the SMART component, similar to the WR method \citep{seewald2020SampleSize}. In a two-stage SMART as Example \ref{ex:marginal:model} where $R$ denotes responder status, the weight is
$W_{\widebar d}^S = \frac{ I\{Z_1 = d_1\} (R + I\{Z_2 = d_2\} (1 - R)) }{ P(Z_1 \mid X_0) P(Z_2 \mid R) }$.
For non-responders, the numerator is 1 since their data align with only one DTR, whereas responders contribute to multiple consistent DTRs (e.g., $(Z_1, Z_2) = (1, 0)$ aligns with $\widebar d = (1, 1)$ and $(1, -1)$).
The MRT weight $W_t^M = \frac{\tilde{p}_t(A_t)}{p_t(A_t \mid H_t)}$ resembles the WCLS approach \citep{boruvka2018assessing}. We set the numerator $\tilde{p}_t(A_t)$ to a constant to target marginal intervention effects, in contrast to WCLS's use of moderator-dependent weights for conditional effects. 
\end{remark}

\paragraph*{Step 2: Estimating $\gamma$}
Having obtained predictions of the expected marginal proximal outcome from Step 1 defined by
\begin{align}
\hat Y_{t+1} & := (a_t - \rho) f_t (\widebar d)^\top \hat \beta + m_t(\widebar d)^\top \hat{\eta},
\end{align}
we next project these predictions onto the space of $m_t (\widebar{d})$ to directly estimate the marginal intervention effects of DTRs averaging over FTS interventions. Instead of regressing on the observed proximal outcome $Y_{t+1}$, we use the predicted outcomes from Step 1 to ensure compatibility with the estimating equation \eqref{eq:ee:beta}.
The second weighted and centered estimating function is
\begin{align}
\begin{split} \label{eq:ee:gamma}
U_2 (\gamma) = 
\sum_{\bar d \in \cD} \sum\limits_{t=1}^T 
W_{\widebar d}^S
\vphantom{\sum\limits_{\widebar z \in \cZ} \sum\limits_{t=1}^T} 
& \left[ \hat Y_{t+1} - m_t (\widebar{d})^\top \gamma \right] 
m_t \left( \widebar{d} \right),
\end{split}
\end{align}
and the estimator $\hat \gamma$ solves $\mathbb{P}_n U_2(\gamma) = 0$.

Denote the true parameters as $\theta^* = (\beta^*, \eta^*, \gamma^*)$. The next proposition states that we can obtain consistent estimators using our joint estimation method. 
\begin{proposition} \label{prop:asymptotic}
Suppose that the causal assumption \ref{assumption:identification} and modeling assumptions \eqref{eq:marginal:model:eq1} hold.  Then
\begin{enumerate}[label = (\arabic*)]
\item \hspace{0.5em} The estimator $(\hat{\beta}, \hat{\eta})$ is consistent and asymptotically normal. Specifically,
\begin{equation}
\sqrt{n} \left(
\begin{pmatrix}
\hat{\beta} \\ \hat{\eta}
\end{pmatrix} - 
\begin{pmatrix}
\beta^* \\ \eta^*
\end{pmatrix} 
\right) \xrightarrow[]{d} \cN \left( 0, \left[ \EE B_1 \right]^{-1} \left[ \EE M_1 \right] \left[ \EE B_1 \right]^{-1} \right),
\end{equation}
where $B_1 = \begin{pmatrix}
\sum_{t=1}^T (A_t - \rho) f_t (\bar d) m_t (\bar d)^\top & \sum_{t=1}^T m_t (\bar d) m_t (\bar d)^\top
\end{pmatrix}^{\otimes 2}$,
and $M_1 = U_1 (\alpha^*, \beta^*, \eta^*)^{\otimes 2}$. A consistent estimator of the asymptotic variance is given by 
$\left[ \PP_n B_1 \right]^{-1} \left[ \PP_n M_1 \right] \left[ \PP_n B_1 \right]^{-1}$.

\item \hspace{0.5em} The estimator $\hat{\gamma}$ is consistent and asymptotically normal. Specifically,
\begin{equation}
\sqrt{n} \left( \hat{\gamma} - \gamma^* \right) \xrightarrow[]{d} \cN \left( 0, \left[ \EE B_2 \right]^{-1} \left[ \EE M_2 \right] \left[ \EE B_2 \right]^{-1} \right),
\end{equation}
where $B_2 = \sum_{t=1}^T m_t (\bar d)^{\otimes 2}$,
$M_2 =  \left( U_2 (\gamma^*) + Q \Omega^{-1} U_1 (\alpha^*, \beta^*, \eta^*) \right)^{\otimes 2}$,
with $$Q = \begin{pmatrix}
0_{r \times p} & \sum_{t=1}^T (A_t - \rho) f_t (\bar d) m_t (\bar d)^\top & \sum_{t=1}^T m_t (\bar d) m_t (\bar d)^\top
\end{pmatrix},$$
$$\Omega = \begin{pmatrix}
\sum_{t=1}^T g_t (H_t) - \mu_{t,\widebar Z_t} (H_t) \\
\sum_{t=1}^T (A_t - \rho) f_t (\bar d) \\
\sum_{t=1}^T m_t (\bar d)
\end{pmatrix}^{\otimes 2}.$$
A consistent estimator of the asymptotic variance is given by 
$ \left[\PP_n B_2 \right]^{-1} \left[ \PP_n M_2 \right] \left[ \PP_n B_2 \right]^{-1}$.
\end{enumerate}
\end{proposition}

\section{Eligibility} \label{sec:method:availability}

In a hybrid SMART-MRT study, both slow-time scale (STS) and fast-time scale (FTS) intervention components restrict randomization to a subset of intervention options for scientific, ethical, or practical considerations.  For the embedded SMART, re-randomization is restricted to different intervention options based on response status.  In the M-Bridge study, for example, non-response was determined through biweekly self-monitoring. Participants who self-identified as a heavy drinker were classified as a heavy drinker. Non-responders had randomization restricted to two bridging strategies.
For the embedded MRT, individuals are often determined to be ineligible to receive an FTS intervention at a particular decision point because intervention delivery is inappropriate, unethical, or unsafe~\citep{klasnja2015microrandomized}.
In the M-Bridge study, whenever a student is flagged as a heavy drinker based on the response to second, third, or fourth self-monitoring survey, the student will transition to Stage 2 at which time they will be considered ``ineligible'' to receive MRT prompts. This restriction implies that the FTS intervention assignment depends on the previously observed proximal outcomes.

Existing analysis of MRT data condition on eligibility to define effects among those available at a particular decision point~\citep{boruvka2018assessing,qian2020EstimatingTimeVarying}.  Since we are primarily interested in synergistic effects, we cannot condition on eligibility as this will lead to causal bias due to conditioning on a post-treatment variable.
We propose to focus on a causal excursion effect that marginalizes over, instead of conditions on, eligibility status.  This is key for the analysis of the M-Bridge study and requires redefining the two FTS intervention options and associated causal estimands as defined in Section~\ref{sec:method:effects}.
Assume that the measurements $X_t$ prior to the $t$-th time point contain the individual's eligibility status, which is denoted by $I_t = 1$ if the individual is eligible and $I_t = 0$ if ineligible.
The potential outcome of eligibility depends on the STS and FTS interventions and can be written as $I_t (\widebar Z_t, \widebar A_{t-1})$. 
In contrast to Section \ref{sec:method:notation}, here we will incorporate decision rule notation to make explicit that MRT treatment $A_t$ is not delivered under ineligibility. 
Define an FTS intervention function as $D(a, i) = ai$ for treatment $a \in \{0, 1\}$, eligibility status $i \in \{0, 1\}$.
Note that $D(a, i)$ equals $1$ when the individual is eligible and treatment is delivered, and equals $0$ otherwise. 
The potential proximal response under a particular DTR $\widebar d$ and MRT treatment sequence $\widebar a_{t}$ is then $Y_{t+1} \left( \widebar d, \left( \widebar{a}_{t-1}, D( a_t, I_t (\widebar d, \widebar a_{t-1}) ) \right) \right)$.

We now incorporate availability into the definition of the marginal proximal effect. 
The marginal proximal effects in \ref{question:interaction:Z} and \ref{question:interaction:A} become:
\begin{enumerate}
\myitem[Q(I.Z.EL)] \hspace{0.5em} \label{question:interaction:Z:availability} (\textbf{I}nteraction effect for $\bZ_t$ marginalized over \textbf{el}igibility) The marginal proximal effect of two DTRs $\widebar d$ versus $\widebar d'$ when a fixed MRT treatment $A_t = a$ is assigned:
\begin{equation} \label{eq:question:interaction:Z:availability}
\EE \left[ Y_{t+1} \left( \widebar d_t, \left( \widebar{a}_{t-1}, D( a, I_t (\widebar d, \widebar a_{t-1}) ) \right) \right) - 
Y_{t+1} \left( \widebar d_t', \left( \widebar{a}_{t-1}, D( a, I_t (\widebar d', \widebar a_{t-1}) ) \right) \right) \right].
\end{equation}

\myitem[Q(I.A.EL)] \hspace{0.5em} \label{question:interaction:A:availability} (\textbf{I}nteraction effect for $\widebar \bA_t$  marginalized over \textbf{el}igibility) The marginal proximal effect of MRT treatments $A_t = 1$ versus $A_t = 0$ when a fixed DTR $\widebar d$ is assigned:
\begin{equation} \label{eq:question:interaction:A:availability}
\EE \left[ Y_{t+1} \left( \widebar d, \left( \widebar{a}_{t-1}, D( 1, I_t (\widebar d, \widebar a_{t-1}) ) \right) \right) - 
Y_{t+1} \left( \widebar d, \left( \widebar{a}_{t-1}, D( 0, I_t (\widebar d, \widebar a_{t-1}) ) \right) \right) \right].
\end{equation}
\end{enumerate}

Assuming consistency, positivity, and sequential ignorability, the marginal proximal outcome can be expressed using observed data similarly to \eqref{eq:marginal:mean:expression:conditional} and \eqref{eq:marginal:mean:expression:weighted}. Estimation requires minor modifications to the proposed estimation strategy in Section~\ref{sec:estimation}.  First, we control for eligibility by setting the auxiliary variable $S_t := I_t$ as in \eqref{eq:ee:beta}. At ineligible decision points, we code $A_t = \rho$ in \eqref{eq:ee:beta} to eliminate the MRT effect term.

\section{Simulation} \label{sec:sim}

We next evaluate our proposed method through extensive simulations motivated from the M-Bridge study.

\subsection{Simulation Setup and Baseline Methods}
We consider a two-stage hybrid SMART-MRT study spanning over $T = 50$ days with the second stage starting on day $t^* = 14$. The embedded SMART follows the restricted design as described in Example \ref{ex:marginal:model}.
The data generation model is based on \cite{boruvka2018assessing} and \cite{seewald2020SampleSize} with minor adjustments made to demonstrate the necessity of our proposed method in a hybrid design.
We assume that we observe a state variable $X_t \in \{-2, 2\}$ whose transition dynamics is given by $p(X_t = 2 \mid A_{t-1}, H_{t-1}) = \mathrm{expit} \left( -A_{t-1} + 0.1 + 0.2 (1-R) \delta_t Z_2 \right)$, where $\delta_t = I\{ t \geq t^*\}$.

Two scenarios with different MRT randomization probability and responder probability are considered. 
In \textbf{scenario I}, we let $p_t (1 \mid H_t) = 0.5$ be a constant, and $P(R = 1 \mid H_t) = 0.6 I\{ Z_1 = 1 \} + 0.45 I\{ Z_1 = -1 \} $ dependent only on $Z_1$. 
In \textbf{scenario II}, we let $p_t (1 \mid H_t)$ be DTR-dependent and $P(R = 1 \mid H_t)$ dependent on the time-varying state $X_t$. See Appendix \ref{sec:sim:setup:additional} for the detailed setup. 
In both scenarios, we assume that individuals are available at all time points.
The proximal outcome is generated as
\begin{align} 
\begin{split} \label{eq:sim:dgp}
Y_{t+1} = &\ 0.5 \tilde{X}_t + 0.1 (A_{t-1} - p_{t-1}(1 \mid H_{t-1})) + \\
& (A_{t} - p_{t}(1 \mid H_{t})) 
(\beta_0^* + \beta_1^* Z_{1} + \beta_2^* Z_{2} + \beta_3^*  \delta_t Z_{1} Z_{2} + \beta_4^* \tilde{X}_t + \beta_5^* \tilde{X}_t Z_{1}) +\\
& \gamma_0^* + \gamma_1^* Z_{1} + \gamma_2^*  \delta_t Z_{2} + \gamma_3^*  \delta_t Z_{1} Z_{2} + \gamma_4^* \tilde{X}_t Z_1 + \gamma_5^* \delta_t (R - p(R = 1 \mid H_t))+ \epsilon_{t}
\end{split}
\end{align}
where $\delta_t = I\{t > t^*\}$, and $\tilde{X}_t = X_{t} - \EE [X_{t} \mid A_{t-1}, H_{t-1}]$ is the centered state.
The residual error $\epsilon_{t}$ follows an AR(1) Gaussian process with $\epsilon_{t} \sim \cN(0, 0.5)$ and $\mathrm{Corr}(\epsilon_{t}, \epsilon_{u}) = 0.5^{|u-t|/2}$. The coefficients are set as 
$$\beta^* = (0.4, -0.3, 0.2, -0.1, 0.4, 0.2) \quad \text{and} \quad \gamma^* = (0, 0.2, -0.1, -0.1, 0.2, 0.2),$$ such that the proximal response depends on the assigned DTR, FTS interventions at the current and previous time points, current state, and their interaction, as well as the responder status.
We generate data with sample sizes $N = 100, 400$, and repeat for 500 replications. 
An analytic expression of the true marginal effects is provided in Appendix \ref{sec:sim:expression}. 

In terms of estimation, the numerator of the MRT weight is $\tilde p_t (1) = \frac{1}{2}$, and the control variables include $X_t$ and $X_t Z_1$. 
The marginal model used in scenario I takes the same form as \eqref{eq:example:model:interaction}.
In scenario II, since the MRT randomization probabilities are different in Stages 1 and 2, we set the marginal model as  
\begin{align}
\begin{split} \label{eq:sim:scenario2:eq1}
E[ Y_{t+1} ((d_1, d_2), \bar A_{t-1}, a))] &= (a - 1/2) \left( \beta_0 + \beta_1 d_1 + \beta_2 \delta_t d_2 + \beta_3 \delta_t d_1 d_2 \right) + \\
&\ \ \ (\eta_0 + \eta_1 d_1) (1-\delta_t) + (\eta_3 + \eta_4 d_1 + \eta_5 d_2 + \eta_6 d_1 d_2) \delta_t.
\end{split} 
\end{align}

For the interaction effects in \ref{question:interaction:Z} and \ref{question:interaction:A}, we only evaluate the performance of the proposed method due to the lack of existing alternatives.
For the average effects, we compare against WCLS for \ref{question:average:D} and WR for \ref{question:average:A}. The detailed setup for the WCLS and WR analyses are described in Appendix \ref{sec:sim:baseline:model}.
In all comparisons, we focus on the average biases, 95\% coverage probabilities (CP), and the relative efficiency (ratio of the asymptotic variance of baseline over the proposed method).


\subsection{Simulation Results} \label{sim:results}
Tables \ref{tab:sim:scenario1:N100} and \ref{tab:sim:scenario2:N100} report the results under sample size $N = 100$ of scenarios I and II, respectively; additional results under $N = 400$ are displayed in Tables \ref{tab:sim:scenario1:N400} and \ref{tab:sim:scenario2:N400} of the Appendix. 
As seen in Table \ref{tab:sim:results:scenario1:N100:contrastsZ:fixedA}, the hybrid method obtains unbiased interaction effects \ref{question:interaction:Z} between two DTRs under a fixed MRT treatment $A_t = a$, in both Stages 1 and 2.
Centering control variables $X_t$ and $X_t Z_1$, by their conditional expectations given the STS interventions as in \eqref{eq:center}, is the key to ensuring unbiased estimates.
The coverage probabilities achieve the nominal 95\% level. 
As for the interaction effects \ref{question:interaction:A} between FTS interventions under a fixed DTR, rows (1) - (6) of Table \ref{tab:sim:results:scenario1:N100:contrastsA} suggest that the hybrid method obtains unbiased estimates and achieves the nominal 95\% coverage probabilities. However, WCLS is subject to biased point estimates because WCLS ignores the restricted embedded SMART in a hybrid design, not accounting for the fact that only non-responders are re-randomized in Stage 2. 
This reason also explains biased estimates and low coverage probabilities obtained by WCLS for the effects \ref{question:average:A} between FTS interventions averaging over DTRs, as shown in rows (7)-(8) of Table \ref{tab:sim:results:scenario1:N100:contrastsA}.
Finally, the estimated effects between DTRs averaging over FTS interventions produced by the hybrid method and WR have nearly zero biases, as demonstrated by Table \ref{tab:sim:results:scenario1:N100:contrastsZAverageA}. The good performance of WR in point estimates is anticipated because the restricted SMART design is taken into consideration by the SMART weights (see \eqref{eq:sim:wr:ee} of the Appendix). 
On the other hand, the hybrid method has higher mean relative efficiency (mRE; the mean standard error of relative efficiency (sdRE) is also displayed). In particular, the hybrid method achieves $4\% \sim 26\%$ efficiency gain compared to WR in scenario I and at most $8\%$ efficiency gain in scenario II. The higher efficiency results from orthogolizing time-varying moderators $X_t$ and $X_t Z_1$ to reduce variance of the estimates.



\begin{table}[t]
\caption{Marginal effect estimation comparisons among three methods in simulation scenario I, where the MRT randomization probability. Sample size $N = 100$.\label{tab:sim:scenario1:N100}}

\begin{subtable}{\linewidth}\centering
    \footnotesize
    \subcaption{\footnotesize{Comparison of effects between $A_t = 1$ and 0, for a fixed DTR or averaging over DTRs.}}\label{tab:sim:results:scenario1:N100:contrastsA}
\begin{tabular}{clllllllll}
\toprule
\multicolumn{1}{c}{ } & \multicolumn{1}{c}{ } & \multicolumn{1}{c}{ } & \multicolumn{1}{c}{ } & \multicolumn{3}{c}{Hybrid} & \multicolumn{3}{c}{WCLS} \\
\cmidrule(l{3pt}r{3pt}){5-7} \cmidrule(l{3pt}r{3pt}){8-10}
 & Stage & Condition & True & Bias & SE & CP & Bias & SE & CP\\
\midrule
1) & 1 & Fix $d_1 = 1$ & 0.1 & 0 & 0.05 & 0.96 & -0.03 & 0.06 & 0.99\\
2) & 1 & Fix $d_1 = -1$ & 0.7 & 0 & 0.05 & 0.97 & 0.06 & 0.05 & 0.75\\
3) & 2 & Fix $\widebar{d} = (1, 1)$ & 0.14 & 0 & 0.06 & 0.97 & -0.01 & 0.07 & 0.97\\
4) & 2 & Fix $\widebar{d} = (1, -1)$ & 0.06 & 0 & 0.06 & 0.97 & -0.04 & 0.11 & 0.98\\
5) & 2 & Fix $\widebar{d} = (-1, 1)$ & 0.86 & 0 & 0.06 & 0.98 & 0.08 & 0.09 & 0.83\\
6) & 2 & Fix $\widebar{d} = (-1, -1)$ & 0.54 & 0 & 0.06 & 0.96 & 0.04 & 0.06 & 0.9\\
\midrule
7) & 1 & Averaging DTR & 0.4 & 0 & 0.03 & 0.97 & 0.02 & 0.04 & 0.92\\
8) & 2 & Averaging DTR & 0.4 & 0 & 0.03 & 0.97 & 0.02 & 0.04 & 0.92\\
\bottomrule
\end{tabular}
\end{subtable}%
\hspace{0.05\textwidth} 

\begin{subtable}{\linewidth}\centering
    \footnotesize
    \subcaption{\footnotesize{Comparison of effects on the proximal outcome between DTRs averaging over FTS interventions.}}\label{tab:sim:results:scenario1:N100:contrastsZAverageA}
\begin{tabular}{clllllllllll}
\toprule
\multicolumn{1}{c}{ } & \multicolumn{1}{c}{ } & \multicolumn{1}{c}{ } & \multicolumn{1}{c}{ } & \multicolumn{3}{c}{Hybrid} & \multicolumn{3}{c}{WR} \\
\cmidrule(l{3pt}r{3pt}){5-7} \cmidrule(l{3pt}r{3pt}){8-10}
 & Stage & Contrast & True & Bias & SE & CP & Bias & SE & CP & mRE & sdRE\\
\midrule
1) & 1 & $d_1 = 1$ vs -1 & 0.4 & 0 & 0.06 & 0.96 & 0 & 0.06 & 0.98 & 1.21 & 0.12\\
\midrule
2) & 2 & $\widebar{d} = (1, 1)$ vs (1, -1) & -0.16 & 0 & 0.07 & 0.95 & 0 & 0.07 & 0.95 & 1.04 & 0.23\\
3) & 2 & $\widebar{d} = (1, 1)$ vs (-1, 1) & 0.32 & 0 & 0.08 & 0.98 & 0 & 0.08 & 0.98 & 1.06 & 0.13\\
4) & 2 & $\widebar{d} = (1, 1)$ vs (-1, -1) & 0.32 & 0 & 0.08 & 0.97 & 0 & 0.08 & 0.97 & 1.10 & 0.14\\
5) & 2 & $\widebar{d} = (1, -1)$ vs (-1, 1) & 0.48 & -0.01 & 0.07 & 0.96 & -0.01 & 0.08 & 0.97 & 1.20 & 0.17\\
6) & 2 & $\widebar{d} = (1, -1)$ vs (-1, -1) & 0.48 & 0 & 0.07 & 0.95 & 0 & 0.08 & 0.97 & 1.26 & 0.18\\
7) & 2 & $\widebar{d} = (-1, 1)$ vs (-1, -1) & 0 & 0.01 & 0.06 & 0.96 & 0.01 & 0.07 & 0.97 & 1.06 & 0.12\\
\bottomrule
\end{tabular}
\end{subtable}%
\hspace{0.05\textwidth} 

\begin{subtable}{\linewidth}\centering
    \footnotesize
    \subcaption{\footnotesize{Comparison of effects on the proximal outcome between DTRs for a fixed fixed MRT treatment. }}\label{tab:sim:results:scenario1:N100:contrastsZ:fixedA}
\begin{tabular}{cllllll}
\toprule
\multicolumn{1}{c}{ } & \multicolumn{1}{c}{ } & \multicolumn{1}{c}{ } & \multicolumn{1}{c}{} & \multicolumn{3}{c}{Hybrid} \\
\cmidrule(l{3pt}r{3pt}){5-7}
 & Stage & Contrast & True & Bias & SE & CP\\
\midrule
1) & 1 & $d_1 = 1$ vs -1 & 0.7 & 0 & 0.06 & 0.98\\
2) & 2 & $\widebar{d} = (1, 1)$ vs (1, -1) & -0.2 & 0 & 0.09 & 0.95\\
3) & 2 & $\widebar{d} = (1, 1)$ vs (-1, 1) & 0.68 & 0 & 0.09 & 0.98\\
4) & 2 & $\widebar{d} = (1, 1)$ vs (-1, -1) & 0.52 & 0.01 & 0.09 & 0.97\\
5) & 2 & $\widebar{d} = (1, -1)$ vs (-1, 1) & 0.88 & -0.01 & 0.08 & 0.97\\
6) & 2 & $\widebar{d} = (1, -1)$ vs (-1, -1) & 0.72 & 0 & 0.08 & 0.97\\
7) & 2 & $\widebar{d} = (-1, 1)$ vs (-1, -1) & -0.16 & 0.01 & 0.08 & 0.95\\
\midrule
8) & 1 & $d_1 = 1$ vs -1 & 0.1 & 0 & 0.07 & 0.96\\
9) & 2 & $\widebar{d} = (1, 1)$ vs (1, -1) & -0.12 & 0 & 0.08 & 0.93\\
10) & 2 & $\widebar{d} = (1, 1)$ vs (-1, 1) & -0.04 & 0 & 0.09 & 0.96\\
11) & 2 & $\widebar{d} = (1, 1)$ vs (-1, -1) & 0.12 & 0 & 0.09 & 0.97\\
12) & 2 & $\widebar{d} = (1, -1)$ vs (-1, 1) & 0.08 & -0.01 & 0.08 & 0.94\\
13) & 2 & $\widebar{d} = (1, -1)$ vs (-1, -1) & 0.24 & 0 & 0.08 & 0.95\\
14) & 2 & $\widebar{d} = (-1, 1)$ vs (-1, -1) & 0.16 & 0 & 0.07 & 0.96\\
\bottomrule
\end{tabular}
\end{subtable}%
\end{table}

\begin{table}[t]
\caption{Marginal effect estimation comparisons among three methods in simulation scenario II, where the MRT randomization probability depends on the DTR assignment. Sample size $N = 100$.\label{tab:sim:scenario2:N100}}

\begin{subtable}{\linewidth}\centering
    \footnotesize
    \subcaption{\footnotesize{Comparison of effects on the proximal outcome between $A_t = 1$ and 0, for a fixed DTR or averaging over DTRs.}}\label{tab:sim:results:scenario2:N100:contrastsA}
\centering
\begin{tabular}{clllllllll}
\toprule
\multicolumn{1}{c}{ } & \multicolumn{1}{c}{ } & \multicolumn{1}{c}{ } & \multicolumn{1}{c}{ } & \multicolumn{3}{c}{Hybrid} & \multicolumn{3}{c}{WCLS} \\
\cmidrule(l{3pt}r{3pt}){5-7} \cmidrule(l{3pt}r{3pt}){8-10}
 & Stage & Condition & True & Bias & SE & CP & Bias & SE & CP\\
\midrule
1) & 1 & Fix $d_1 = 1$ & 0.1 & 0 & 0.05 & 0.94 & -0.02 & 0.07 & 0.92\\
2) & 1 & Fix $d_1 = -1$ & 0.7 & 0 & 0.05 & 0.94 & 0.06 & 0.05 & 0.81\\
3) & 2 & Fix $\widebar{d} = (1, 1)$ & 0.15 & 0 & 0.07 & 0.95 & -0.01 & 0.07 & 0.93\\
4) & 2 & Fix $\widebar{d} = (1, -1)$ & 0.05 & 0 & 0.08 & 0.93 & -0.03 & 0.13 & 0.9\\
5) & 2 & Fix $\widebar{d} = (-1, 1)$ & 0.89 & 0 & 0.08 & 0.96 & 0.07 & 0.1 & 0.86\\
6) & 2 & Fix $\widebar{d} = (-1, -1)$ & 0.51 & 0 & 0.07 & 0.93 & 0.04 & 0.07 & 0.91\\
\midrule
7) & 1 & Averaging DTR & 0.4 & 0 & 0.04 & 0.95 & 0.02 & 0.04 & 0.92\\
8) & 2 & Averaging DTR & 0.4 & 0 & 0.04 & 0.95 & 0.02 & 0.04 & 0.92\\
\bottomrule
\end{tabular}
\end{subtable}%
\hspace{0.05\textwidth} 

\begin{subtable}{\linewidth}\centering
    \footnotesize
    \subcaption{\footnotesize{Comparison of effects on the proximal outcome between DTRs averaging over FTS interventions.}}\label{tab:sim:results:scenario2:N100:contrastsZAverageA}
\centering
\begin{tabular}{clllllllllll}
\toprule
\multicolumn{1}{c}{ } & \multicolumn{1}{c}{ } & \multicolumn{1}{c}{ } & \multicolumn{1}{c}{ } & \multicolumn{3}{c}{Hybrid} & \multicolumn{3}{c}{WR} \\
\cmidrule(l{3pt}r{3pt}){5-7} \cmidrule(l{3pt}r{3pt}){8-10}
 & Stage & Contrast & True & Bias & SE & CP & Bias & SE & CP & mRE & sdRE\\
\midrule
1) & 1 & $d_1 = 1$ vs -1 & 0.4 & 0 & 0.06 & 0.95 & 0 & 0.06 & 0.95 & 1 & 0\\
\midrule
2) & 2 & $\widebar{d} = (1, 1)$ vs (1, -1) & -0.2 & 0 & 0.08 & 0.95 & 0 & 0.08 & 0.95 & 1 & 0\\
3) & 2 & $\widebar{d} = (1, 1)$ vs (-1, 1) & 0.3 & 0 & 0.08 & 0.95 & 0 & 0.08 & 0.95 & 1 & 0\\
4) & 2 & $\widebar{d} = (1, 1)$ vs (-1, -1) & 0.3 & 0 & 0.08 & 0.96 & 0 & 0.08 & 0.96 & 1 & 0\\
5) & 2 & $\widebar{d} = (1, -1)$ vs (-1, 1) & 0.5 & 0 & 0.08 & 0.92 & 0 & 0.08 & 0.92 & 1 & 0\\
6) & 2 & $\widebar{d} = (1, -1)$ vs (-1, -1) & 0.5 & 0 & 0.08 & 0.94 & 0 & 0.08 & 0.94 & 1 & 0\\
7) & 2 & $\widebar{d} = (-1, 1)$ vs (-1, -1) & 0 & 0 & 0.07 & 0.93 & 0 & 0.07 & 0.93 & 1 & 0\\
\bottomrule
\end{tabular}
\end{subtable}%
\hspace{0.05\textwidth} 

\begin{subtable}{\linewidth}\centering
    \footnotesize
    \subcaption{ \footnotesize{Comparison of effects on the proximal outcome between DTRs for a fixed MRT treatment.}}\label{tab:sim:results:scenario2:N100:contrastsZ:fixedA}
\begin{tabular}{cllllll}
\toprule
\multicolumn{1}{c}{ } & \multicolumn{1}{c}{ } & \multicolumn{1}{c}{ } & \multicolumn{1}{c}{} & \multicolumn{3}{c}{Hybrid} \\
\cmidrule(l{3pt}r{3pt}){5-7}
 & Stage & Contrast & True & Bias & SE & CP\\
\midrule
1) & 1 & $d_1 = 1$ vs -1 & 0.62 & 0 & 0.11 & 0.91\\
2) & 2 & $\widebar{d} = (1, 1)$ vs (1, -1) & -0.24 & 0 & 0.11 & 0.96\\
3) & 2 & $\widebar{d} = (1, 1)$ vs (-1, 1) & 0.46 & 0 & 0.09 & 0.94\\
4) & 2 & $\widebar{d} = (1, 1)$ vs (-1, -1) & 0.48 & 0 & 0.1 & 0.95\\
5) & 2 & $\widebar{d} = (1, -1)$ vs (-1, 1) & 0.7 & -0.01 & 0.1 & 0.93\\
6) & 2 & $\widebar{d} = (1, -1)$ vs (-1, -1) & 0.73 & 0 & 0.11 & 0.94\\
7) & 2 & $\widebar{d} = (-1, 1)$ vs (-1, -1) & 0.03 & 0 & 0.08 & 0.94\\
\midrule
8) & 1 & $d_1 = 1$ vs -1 & 0.02 & 0 & 0.1 & 0.89\\
9) & 2 & $\widebar{d} = (1, 1)$ vs (1, -1) & -0.14 & 0 & 0.09 & 0.93\\
10) & 2 & $\widebar{d} = (1, 1)$ vs (-1, 1) & -0.28 & 0 & 0.11 & 0.92\\
11) & 2 & $\widebar{d} = (1, 1)$ vs (-1, -1) & 0.12 & 0 & 0.1 & 0.93\\
12) & 2 & $\widebar{d} = (1, -1)$ vs (-1, 1) & -0.14 & 0 & 0.1 & 0.91\\
13) & 2 & $\widebar{d} = (1, -1)$ vs (-1, -1) & 0.27 & 0 & 0.09 & 0.92\\
14) & 2 & $\widebar{d} = (-1, 1)$ vs (-1, -1) & 0.4 & 0 & 0.1 & 0.96\\
\bottomrule
\end{tabular}
\end{subtable}%
\end{table}

\section{Application to M-Bridge Study} \label{sec:data}


\subsection{Study Design and Questions}
In this section, we focus on a subset of $N = 428$ students from the M-Bridge study. Details of dataset construction as part of the larger M-Bridge study are provided in Appendix \ref{sec:data:construction}.
Recall that the M-Bridge study employs a SMART-MRT hybrid design where the SMART involved two stages of randomization -- first-stage random assignment (1:1 ratio) early or late timing of an initial web-based intervention using personalized normative feedback (PNF), with non-responders being re-randomized (1:1 ratio) to an email with ineligible alcohol use intervention resources (Email), or an invitation to interact with an online health coach (Coach). Non-response was determined based on identifying as a heavy drinker via bi-weekly self-monitoring and therefore can occur at weeks 2, 4, 6, or 8. Those not identified as heavy drinkers (i.e., responders) continued with self-monitoring alone. The MRT involved bi-weekly randomization of those in the self-monitoring condition to two types of prompts (1:1 ratio) encouraging participants to self-monitor their alcohol use: either a self-interest (SI) prompt or a pro-social (PS) prompt.

Since not all students were randomized to a prompt at every time point, we adopt the eligibility framework in Section \ref{sec:method:availability}. A student was ineligible ($I_t = 0, t = 2, 3, 4$) to receive an MRT intervention $A_t$ if the student was identified as a heavy drinker at any time point before $t$; otherwise $I_t = 1$. 
We point out that different from Example \ref{ex:marginal:model} where the MRT spans both stages, the M-Bridge only involves MRT in Stage 1. 
We are interested in (a) the main and interaction between MRT and SMART intervention effects in Stage 1, and (b) the main SMART intervention effects in Stage 2 (since MRT does not span Stage 2), averaging over the study population. The associated estimands are outlined in Table \ref{tab:data:estimand}.
We let are $f_t(\widebar d) = (1, d_1, \tilde t, \tilde t d_1)$ and $m_t(\widebar d) = (1, d_1, \delta_t d_2, \delta_t d_1 d_2)$,
where $\tilde t$ denotes the week of SM surveys centered by mean and scaled by standard deviation, and $\delta_t = I\{ t \geq 4\}$ is an indicator of whether a student is in Stage 2. 
In estimation, we set $\tilde{p}_t = 0.5$.
The control variables include baseline sex (female/male) and whether the student pledges Greek life.

\begin{table}[ht]
\centering
\footnotesize
\caption{Causal estimands for the M-Bridge study}
\label{tab:data:estimand}
\begin{tabularx}{\linewidth}{XX}
\toprule
Scientific Question & Causal Estimand \\
\midrule
(1) What is the effect on maximum number of drinks for an SI prompt compared to a PS prompt in Stage 1, given a fixed initial timing of PNF $d_1 \in \{-1, 1\}$? &
\vspace{-0.4em} 
$\begin{aligned}
\EE &\left[ Y_{t+1} \left( d_1, \left( \widebar{a}_{t-1}, D( 1, I_t (d_1, \widebar a_{t-1}) ) \right) \right) \right. \\
& - \left.
Y_{t+1} \left( d_1, \left( \widebar{a}_{t-1}, D( 0, I_t (d_1, \widebar a_{t-1}) ) \right) \right) \middle\vert X_0 \right], \\
& t = 2, 3, 4.
\end{aligned}$ \\
\hline
(2) What is the effect on maximum number of drinks for early compared to late timing of PNF in Stage 1, given a fixed prompt type $A_t = a$? & 
\vspace{-0.4em} 
$\begin{aligned}
\EE &\left[ Y_{t+1} \left( 1, \left( \widebar{a}_{t-1}, D( a, I_t (1, \widebar a_{t-1}) ) \right) \right) \right. \\
& - \left.
Y_{t+1} \left( -1, \left( \widebar{a}_{t-1}, D( a, I_t (-1, \widebar a_{t-1}) ) \right) \right) \middle\vert X_0 \right], \\
& t = 2, 3, 4.
\end{aligned}$ \\
\hline 
(3) What is the effect on maximum number of drinks for an SI prompt compared to a PS prompt in Stage 1, averaging over initial timings of PNF? & 
\vspace{-0.4em} 
$\begin{aligned}
&\sum_{l_1 \in \{-1, 1\}} P(d_1 = l_1)\EE \left[ Y_{t+1} \left( l_1, \left( \widebar{a}_{t-1}, D( 1, I_t (l_1, \widebar a_{t-1}) ) \right) \right) \right. \\
& - \left.
Y_{t+1} \left( l_1, \left( \widebar{a}_{t-1}, D( 0, I_t (l_1, \widebar a_{t-1}) ) \right) \right) \middle\vert X_0 \right], t = 2, 3, 4.
\end{aligned}$ \\
\hline
(4) What is the effect on maximum number of drinks for early compared to late timing of PNF, averaging over prompt types? &
\vspace{-0.4em} 
$\begin{aligned}
\EE &\left[ Y_{t+1} \left( 1, \left( \widebar{A}_{t-1}, D( A_t, I_t (1, \widebar A_{t-1}) ) \right) \right) \right. \\
& - \left.
Y_{t+1} \left( -1, \left( \widebar{A}_{t-1}, D( A_t, I_t (-1, \widebar A_{t-1}) ) \right) \right) \middle\vert X_0 \right].
\end{aligned}$ \\
\bottomrule
\end{tabularx}
\end{table}

\subsection{Results} \label{sec:data:result}

\begin{figure}[ht]
    \centering
    \begin{minipage}[t]{0.51\textwidth}
        \centering
        \includegraphics[width=\textwidth]{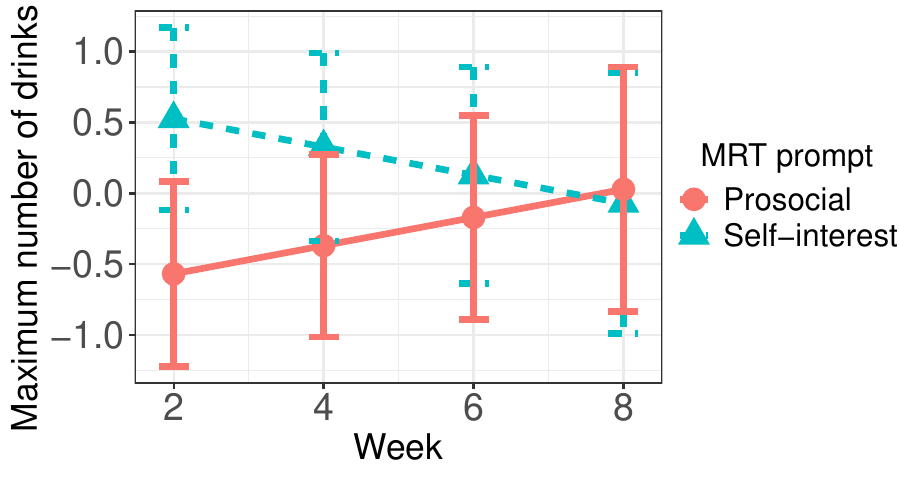}
        \vspace{-2em}
        \caption*{(a) Effect of Early vs Late timing of PNF given fixed bridging strategy.}
        \label{fig:line_constrast_Z1_fixedA}
    \end{minipage}
    \hfill
     \begin{minipage}[t]{0.46\textwidth}
        \centering
        \includegraphics[width=\textwidth]{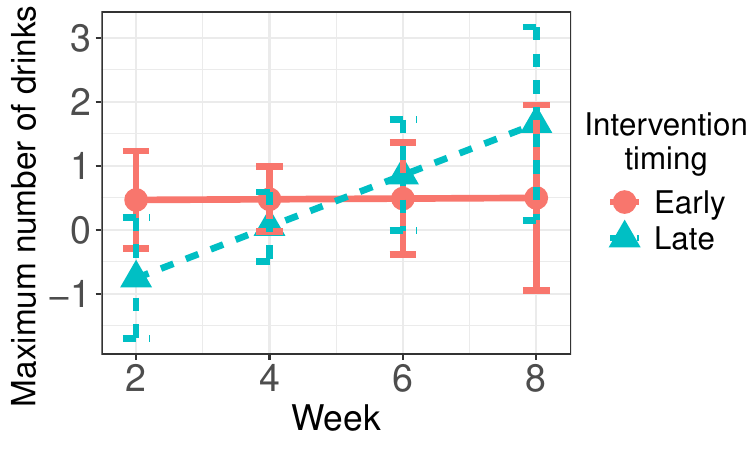}
        \vspace{-2em}
        \caption*{(b) Effect of SI vs. PS prompt given fixed initial timing of PNF.}
        \label{fig:line_constrast_A_fixedZ1}
    \end{minipage}
    \begin{minipage}{0.5\textwidth}
        \centering
        \includegraphics[width=\textwidth]{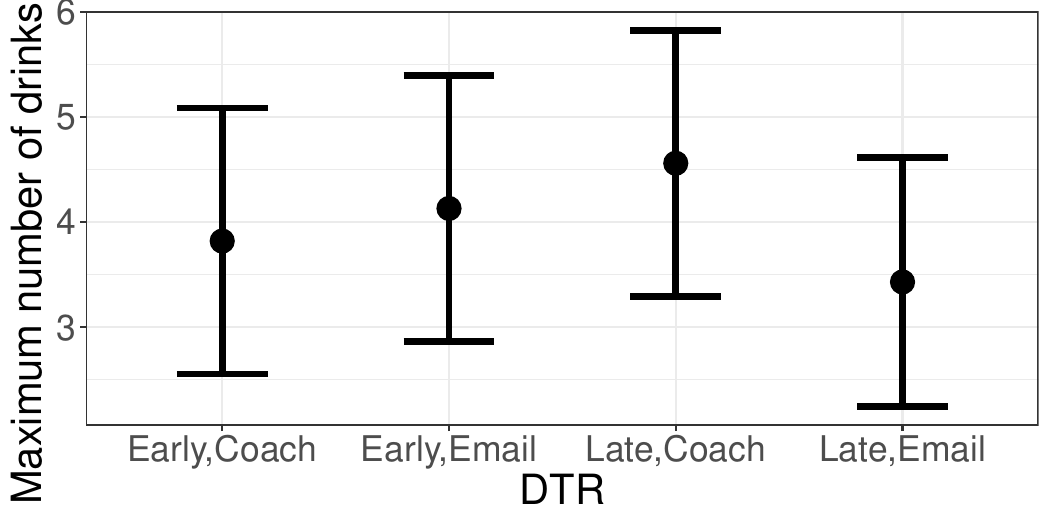}
        \vspace{-2em}
        \label{fig:line_constrast_Z}
    \end{minipage}
    \begin{minipage}{0.45\textwidth}
        \small{(c) Marginal mean outcome for a fixed universal intervention timing combined with bridging strategy.}
    \end{minipage}
    \caption{Estimated trajectories in the M-Bridge study, where the proximal outcome is the maximum number of alcoholic drinks a student consumed within a 24-hour period over the past two weeks.}
    \label{fig:data:trajectories}
\end{figure}

The estimated treatment effects are displayed in Figure \ref{fig:data:trajectories}, and coefficient estimates are reported in Table \ref{tab:data:coefficients}. 
Our first question was ``What is the effect of Early versus Late PNF timing on maximum number of drinks when fixing the MRT intervention to either PS or SI prompt.  At week 2, Figure~\ref{fig:data:trajectories}(a) shows that the effect at week 2 is positive when fixing to SI prompt and negative when fixing to PS prompt.  This suggests the benefit of Early PNF timing relative to Late PNF timing may be synergistic with PS prompts at least initially. 
On the other hand, the effect is anti-synergistic with SI prompts in Stage 1, i.e., weeks 2 and 4 effects are positive.  Both of these effects diminish to zero by week 8.
Our second question was ``What is the effect on maximum number of drinks for an SI prompt compared to a PS prompt in Stage 1, given a fixed initial timing of PNF?''  Figure~\ref{fig:data:trajectories}(b) shows that the effects when PNF timing is fixed to Early is negligible and does not vary over weeks. When PNF timing is fixed to Late, the SI prompts lead to reduction in the maximum number of drinks reported relative to PS prompts during week 2, but that the effect changes sign by week 8.  
Our final question was ``What is the effect on the maximum number of alcoholic drinks of the four embedded DTRs, averaging over the MRT intervention component?''  Figure~\ref{fig:data:trajectories}(c) presents the estimated marginal mean outcome for each of the four DTRs.  The DTR that starts with Late PNF and provides Coaching for non-responders led to the highest marginal mean outcome of 4.5 (95\% CI: 3.30 - 5.83), while the DTR that starts with Late PNF and provides a resource email for non-responders was associated with the lowest marginal mean outcome o 3.4 (95\% CI: 2.25--4.62). 




\begin{table}[ht]
\caption{Estimated coefficients and 95\% confidence intervals in M-Bridge data analysis} 
\label{tab:data:coefficients}
\centering
\begin{tabular}{llrl}
  \hline
Parameter & Variable & Estimate & 95\% CI \\ 
  \hline
$\beta_0$ & $(a_t - 0.5)$ & 0.47 & (0.36, 0.57) \\ 
  $\beta_1$ & $(a_t - 0.5)d_1$ & 0.02 & (-0.09, 0.12) \\ 
  $\beta_2$ & $(a_t - 0.5)t$ & 0.46 & (0.3, 0.61) \\ 
  $\beta_3$ & $(a_t - 0.5)d_1t$ & -0.44 & (-0.6, -0.29) \\ 
  $\gamma_0$ & Intercept & 3.99 & (3.96, 4.01) \\ 
  $\gamma_1$ & $d_1$ & -0.01 & (-0.04, 0.02) \\ 
  $\gamma_2$ & $\delta_t d_2$ & 0.20 & (-0.17, 0.57) \\ 
  $\gamma_3$ & $d_1\delta_t d_2$ & -0.36 & (-0.73, 0.01) \\ 
   \hline
\end{tabular}
\end{table}

\section{Discussion} \label{sec:discussion}

In this paper, we proposed a novel statistical framework for analyzing data from hybrid SMART-MRT studies, enabling the joint estimation of the synergistic and marginal causal effects of interventions operating at multiple timescales. Our method integrates information from both SMART and MRT components in a hybrid SMART-MRT, allowing researchers to assess how long-term adaptive intervention strategies interact with just-in-time support. In the the M-Bridge study, our method captures the combined effects of preventive strategies and JITAIs on reducing binge drinking among first-year college students. This approach provides a more comprehensive understanding of intervention effectiveness than existing methods, which typically analyze SMART and MRT data in isolation.

Our work has several limitations. First, our analysis focuses on marginal effects, providing population-level estimates of intervention impact rather than conditional effects that account for individual heterogeneity. Future research is needed to develop methods that accommodate personalized treatment effects and examine mediation mechanisms.
Second, we did not explicitly handle missing data, except for cases related to participant availability in the MRT component. In practice, missingness may occur due to loss to follow-up, incomplete survey responses, or disengagement from intervention components. Future work should explore robust imputation strategies and inverse probability weighting methods to address missingness, particularly in longitudinal hybrid trial settings where data sparsity can be an issue.

\section*{Data Availability and Code}

\sloppy 
The data that support the findings of this study are protected under a Data and Materials Distribution Agreement (DMDA). Access to the application data is available upon request.
Code to reproduce simulations is available online at
\href{http://github.com/limengbinggz/ddtlcm}{\nolinkurl{http://github.com/limengbinggz/ddtlcm}}.




\appendix
\section{Notation Summary}
Table~\ref{tab:notation} summarizes all relevant notation used in defining the data, estimands, and estimators for SMART-MRT hybrid designs.

\begin{table}[ht]
\centering
\caption{Notation Summary for Hybrid SMART-MRT Design}
\label{tab:notation}
\begin{tabularx}{\linewidth}{lX}
\toprule
Notation & Description \\
\midrule
$t^*$ & Time of entering Stage 2. \\
$X_0$ & Pre-treatment baseline covariates. \\
$X_t$ & Contextual and individual information at time $t$. \\
$Z_1, Z_2$ & SMART treatments in Stages 1 and 2, respectively. \\
$\widebar{Z}_t$ & Sequence of SMART treatments up to time $t$ ($\widebar{Z}_t = Z_1$ for $t < t^*$, $\widebar{Z}_t = (Z_1, Z_2)$ for $t \geq t^*$). \\
$A_t$ & MRT treatment at time $t$. \\
$Y_{t+1}$ & Proximal response subsequent to treatment $A_t$. \\
$\widebar{A}_t, \widebar{X}_t, \widebar{Y}_t$ & Sequence of MRT treatments, contextual information, and proximal responses up to time $t$, respectively. \\
$H_t = \left( \widebar{X}_t, \widebar{Z}_t, \widebar{A}_{t}, \widebar{Y}_t \right)$ & Complete history of observed data up to time $t$. \\
$Y_{t+1} \left( \widebar{z}_t, \widebar{a}_t \right)$ & Potential outcome for proximal response under specific SMART and MRT treatment sequences. \\
$R$ & Tailoring variable for Stage 2 SMART treatment (e.g., responder status). \\
$\widebar{d} = (d_1, d_2)$ & Dynamic treatment regime (DTR), where $d_1$ and $d_2$ are decision rules for SMART treatments in Stages 1 and 2. \\
$f_t (\widebar d)$ & $p$-dimensional vector functions of DTRs $\widebar d$ and baseline variables $X_0$ \\
$m_t (\widebar d)$ & $q$-dimensional vector functions of DTRs $\widebar d$ and baseline variables $X_0$ \\
$S_t$ & $r$-dimensional time-varying effect moderator \\
$\psi_t (\widebar d)$ & $r$-dimensional centering function of $S_t$ \\
\bottomrule
\end{tabularx}
\end{table}

\section{Identification Results} \label{sec:proof:observeddata}

For simplicity, we omit the baseline variables $X_0$ from the conditional expectations.

We first prove the first equation \eqref{eq:marginal:mean:expression:conditional}.

For simplicity, we denote the history prior to the first SMART treatment assignment as $H_0 = (X_0, X_1)$. By sequential ignorability SI(a) in Assuption \ref{assumption:identification}, we have 
\begin{align*}
  \EE \left[ Y_{t+1} \left( \widebar d_t, (\widebar{A}_{t-1}, a) \right) \middle\vert X_0 \right]
  &= \EE \left[ Y_{t+1} \left( \widebar d_t, (\widebar{A}_{t-1}, a) \right) \middle\vert H_0 \right] \\
  &= \EE \left[ \EE \left[ Y_{t+1} \left( \widebar d_t, (\widebar{A}_{t-1}, a) \right) \middle\vert H_0, Z_1 = d_1 (H_0) \right] \right].
\end{align*} 

Recall that by consistency, $H_{t^*} (Z_1, \widebar{A}_{t^*-1}) = H_{t^*}$ and $H_{t} (\widebar Z, \widebar{A}_{t-1})  = H_t$.
\begin{align}
&\ \EE \left[ \EE \left[ Y_{t+1} \left( \widebar d, (\widebar{A}_{t-1}, a) \right) \middle\vert H_0, Z_1 = d_1 (H_0) \right] \right] \nonumber \\
=&\ \EE \left\{ \EE \left[ \EE \left[ Y_{t+1} \left( \widebar d, (\widebar{A}_{t-1}, a) \right) 
\middle\vert H_{t^*-1} (Z_1, \widebar{A}_{t^*-1}) \right] 
\middle\vert H_0, Z_1 = d_1 (H_0) \right] \right\} \nonumber \\
=&\ \EE \left\{ \EE \left[ \EE \left[ Y_{t+1} \left( \widebar d, (\widebar{A}_{t-1}, a) \right) 
\middle\vert H_{t^*-1} \right] 
\middle\vert H_0, Z_1 = d_1 (H_0) \right] \right\} \nonumber \\
=&\ \EE \left\{ \EE \left[ \EE \left[ Y_{t+1} \left( \widebar d_t, (\widebar{A}_{t-1}, a) \right) 
\middle\vert H_{t^*-1}, Z_2 = d_2 (H_{t^*-1})\right] 
\middle\vert H_0, Z_1 = d_1 (H_0) \right] \right\} \nonumber \\
=&\ \EE \left\{ \EE \left\{ \EE \left[ \EE 
\left[ Y_{t+1} \left( \widebar d, (\widebar{A}_{t-1}, a) \right) \middle\vert H_{t} (\widebar Z, \widebar{A}_{t-1}) \right] 
\middle\vert H_{t^*-1}, Z_2 = d_2 (H_{t^*-1})\right] 
\middle\vert H_0, Z_1 = d_1 (H_0) \right\} \right\} \nonumber \\
=&\ \EE \left\{ \EE \left\{ \EE \left[ \EE 
\left[ Y_{t+1} \left( \widebar d, (\widebar{A}_{t-1}, a) \right) \middle\vert H_{t} \right] 
\middle\vert H_{t^*}, Z_2 = d_2 (H_{t^*-1})\right] 
\middle\vert H_0, Z_1 = d_1 (H_0) \right\} \right\}, \nonumber
\end{align}
where the first and third equations follow from sequential ignorability SI(b), and the second and fourth equations follow from consistency.
Following the same reasoning, the last equation becomes
\begin{align}
&\ \EE \left\{ \EE \left\{ \EE \left[ \EE 
\left[ Y_{t+1} \left( \widebar d, (\widebar{A}_{t-1}, a) \right) \middle\vert H_{t}, A_t \right] 
\middle\vert H_{t^*}, Z_2 = d_2 (H_{t^*-1})\right] 
\middle\vert H_0, Z_1 = d_1 (H_0) \right\} \right\} \nonumber \\
=&\ \EE \left\{ \EE \left\{ \EE \left[ \EE 
\left[ Y_{t+1} \left( \widebar d, (\widebar{A}_{t-1}, A_t) \right) \middle\vert H_{t}, A_t = a \right] 
\middle\vert H_{t^*}, Z_2 = d_2 (H_{t^*-1})\right] 
\middle\vert H_0, Z_1 = d_1 (H_0) \right\} \right\} \nonumber \\
=&\ \EE \left\{ \EE \left\{ \EE \left[ \EE 
\left[ Y_{t+1} \middle\vert H_{t}, A_t = a \right] 
\middle\vert H_{t^*}, Z_2 = d_2 (H_{t^*-1})\right] 
\middle\vert H_0, Z_1 = d_1 (H_0) \right\} \right\}, \nonumber
\end{align}
following SI(b) and consistency.

\sloppy Next, we show \eqref{eq:marginal:mean:expression:weighted}. For simplicity, we will omit the baseline variable $X_0$ from $\EE \left[ Y_{t+1} \left( \widebar d, (\widebar{A}_{t-1}, a) \right) \middle\vert X_0 \right]$, only focusing on $\EE \left[ Y_{t+1} \left( \widebar d, (\widebar{A}_{t-1}, a) \right) \right]$.
By sequential ignorability, we know that 
$
 \EE \left[ 
    Y_{t+1} \left( \widebar z, (\widebar{A}_{t-1}, a) \right)
  1 \{z_1 = d_1 (H_0) \} 1 \{z_2 = d_2 (H_{t^*-1}) \} 
  1 \{A_t = a \} \middle\vert H_t \right] 
$ equals
$\EE \left[ 
    Y_{t+1} \left( \widebar z, (\widebar{A}_{t-1}, a) \right) \middle\vert H_t \right]
  \EE \left[ 1 \{z_1 = d_1 (H_0)\} 1 \{z_2 = d_2 (H_{t^*-1}) \} 1 \{A_t = a \} \middle\vert H_t \right]
$.
We then have
\begin{align*}
  &\ \EE \left[ Y_{t+1} \left( \widebar d, (\widebar{A}_{t-1}, a) \right) \right] \nonumber \\
  =&\ \sum_{z_1 \in \cZ_1} \sum_{z_2 \in \cZ_2} \EE \left[ Y_{t+1} \left( \widebar z, (\widebar{A}_{t-1}, a) \right) 
  1\{z_1 = d_1 (X_0)\} 1\{z_2 = d_2 (H_{t^*-1})\} 
  \right] \nonumber \\
  =&\ \sum_{z_1 \in \cZ_1} \sum_{z_2 \in \cZ_2} \EE \left[ \EE \left[ 
    Y_{t+1} \left( \widebar z, (\widebar{A}_{t-1}, a) \right) 
    1\{z_1 = d_1 (X_0)\} 1\{z_2 = d_2 (H_{t^*-1})\} \middle\vert H_t \right]\right] \nonumber \\
  =&\ \sum_{z_1 \in \cZ_1} \sum_{z_2 \in \cZ_2} \EE \left[ 
    \EE \left[ Y_{t+1} \left( \widebar z, (\widebar{A}_{t-1}, a) \right) \middle\vert H_t \right] \phantom{\frac{ 1 \{z_2 = d_2 (H_{t^*-1}) \} }{ P(\widebar Z = \widebar z \mid H_t) }}\right. \\
    & \hspace{6em} \left. \cdot \frac{\EE \left[ 1 \{z_1 = d_1 (H_0) \} 1 \{z_2 = d_2 (H_{t^*-1}) \} \middle\vert H_t \right] 1 \{A_t = a \} }{ P(\widebar Z = \widebar z \mid H_t) p_t(a \mid H_t,\widebar Z = \widebar z)}
    \right] \nonumber \\
  =& \ \mathbb{E} \left[ \mathbb{E} \left[ 
  \sum_{z_1 \in \cZ_1} \sum_{z_2 \in \cZ_2} 
  \frac{ 1 \{z_1 = d_1 (X_1) \} 1 \{z_2 = d_2 (H_{t^*-1}) \} }{P(Z_1 = z_1 \mid X_1) P(Z_2 = z_2 \mid Z_1 = z_1, H_{t^*-1})}
  \frac{ 1 \{A_t = a \}}{ p_t(a \mid H_t,\widebar Z = \widebar z) }
  Y \middle\vert H_t \right] \right] \\
  =& \ \mathbb{E} \left[ 
  \sum_{z_1 \in \cZ_1} \sum_{z_2 \in \cZ_2} 
  \frac{ 1 \{z_1 = d_1 (X_1) \} 1 \{z_2 = d_2 (H_{t^*-1}) \} }{P(Z_1 = z_1 \mid X_1) P(Z_2 = z_2 \mid Z_1 = z_1, H_{t^*-1})}
  \frac{ 1 \{A_t = a \}}{ p_t(a \mid H_t,\widebar Z = \widebar z) }
  Y \right]. 
\end{align*}


\section{Rationale for \ref{prop:asymptotic}} \label{sec:marginal:model:eq2}
From Example \ref{ex:marginal:model}, we can see that $\beta$ in \eqref{eq:marginal:model:eq1} can be interpreted as the MRT effect given a fixed DTR. On the other hand, $\eta$ has a more complicated interpretation that it embeds the comparison of proximal effects between DTRs averaging over MRT treatments, as if the MRT randomization probability were $\rho$ for $A_t = 1$. To see this, note that 
\begin{align*}    
& \EE_{A_t \sim \rho} \left[ Y_{t+1} \left( \widebar d_t, (\widebar{A}_{t-1}, A_t) \right) \middle\vert X_0 \right] \\
=&\ \EE_{A_t \sim \rho} \left\{  (A_t - \rho) f_t (\widebar d_t, X_0)^\top \beta + m_t(\widebar d_t, X_0)^\top \eta \right\} \\
=&\ \rho \left[ (1 - \rho) f_t (\widebar d_t, X_0)^\top \beta + m_t(\widebar d_t, X_0)^\top \eta \right] + 
(1- \rho) \left[ (0 - \rho) f_t (\widebar d_t, X_0)^\top \beta + m_t(\widebar d_t, X_0)^\top \eta \right] \\
=&\ m_t(\widebar d_t, X_0)^\top \eta.
\end{align*}
Hence, 
$\EE_{A_t \sim \rho} \left[ Y_{t+1} \left( \widebar d_t, (\widebar{A}_{t-1}, A_t) \right) - Y_{t+1} \left( \widebar d_t', (\widebar{A}_{t-1}, A_t) \right) \middle\vert X_0 \right] = \left( m_t(\widebar d_t, X_0) - m_t(\widebar d_t', X_0) \right)^\top \eta$,
which is a different quantity than what the scientific question \ref{question:average:D} is concerned with. However, we point out that we only care about the marginal effects comparing MRT treatments or DTRs, which are linear combinations of the coefficient $(\beta, \eta)$, rather than the coefficient itself.
In addition, Let $H_t^0 (\widebar d_t, \widebar{A}_{t-1}) \subseteq H_{t} (\widebar d_t, \widebar{A}_{t-1})$ denote a subset of the potential history which MRT randomization probability depends on. Note that by assumption \ref{assumption:identification}, the MRT randomization probability $P(A_t = a \mid H_{t}^0 (\widebar d_t, \widebar{A}_{t-1})) = P(A_t = a \mid H_{t}^0) =: p_t (a \mid H_{t}^0)$.
Then \eqref{eq:question:average:D} corresponds to 
\begin{align}    
& \EE \left[ Y_{t+1} \left( \widebar d_t, (\widebar{A}_{t-1}, A_t) \right) \middle\vert X_0 \right]  \nonumber \\
=&\ \int_{H_t^0} \sum_{a=0}^1 \EE \left[ 
Y_{t+1} \left( \widebar d_t, (\widebar{A}_{t-1}, a) \right) \middle\vert X_0 \right] 
P(a \mid H_{t}^0 (\widebar d_t, \widebar{A}_{t-1}))
d P( H_{t}^0 (\widebar d_t, \widebar{A}_{t-1}) ) \\
=&\ \int_{H_t^0} \sum_{a=0}^1 \left[ 
(a - \rho) f_t (\widebar d_t, X_0)^\top \beta + m_t(\widebar d_t, X_0)^\top \eta \right] 
p_t (a \mid H_{t}^0)
d P( H_{t}^0 ). \label{eq:expectation:average:D} 
\end{align}
Although $p_t (a \mid H_{t}^0)$ is known in a hybrid design study, the integral requires knowledge about the distribution of the potential history. This creates difficulty in evaluating \eqref{eq:question:average:D} directly. On the other hand, notice that \eqref{eq:expectation:average:D} can be further written as
\begin{align}    
\EE \left[ Y_{t+1} \left( \widebar d_t, (\widebar{A}_{t-1}, A_t) \right) \middle\vert X_0 \right] &=
\begin{pmatrix}
\int_{H_t^0} \sum_{a=0}^1 
(a - \rho) f_t (\widebar d_t, X_0) p_t (a \mid H_{t}^0)
d P( H_{t}^0 ) \\
\int_{H_t^0} \sum_{a=0}^1 
m_t(\widebar d_t, X_0) p_t (a \mid H_{t}^0)
d P( H_{t}^0 ) 
\end{pmatrix}^\top
\begin{pmatrix}
\beta \\ \eta
\end{pmatrix} \label{eq:expectation:average:D:integral} \\
&=: m_t(\widebar d_t, X_0)^\top \gamma, \label{eq:marginal:model:eq2}
\end{align}
where $m_t(\widebar d_t, X_0)$ is an $r$-dimensional function of $\widebar d_t$ and $X_0$ that combines the shared terms in the design matrices in \eqref{eq:expectation:average:D:integral}, and $\gamma$ is a vector of coefficients. This allows us to directly model the expected potential proximal outcome of a DTR averaging over MRT treatment sequences. The coefficient $\gamma$ has the interpretation of the comparison of proximal effects between DTRs averaging over MRT treatments, under the actual MRT randomization probability by study design. This is because $\EE \left[ Y_{t+1} \left( \widebar d_t, (\widebar{A}_{t-1}, A_t) \right) - Y_{t+1} \left( \widebar d_t', (\widebar{A}_{t-1}, A_t) \right) \middle\vert X_0 \right] = \left[ m_t(\widebar d_t, X_0) - m_t(\widebar d_t', X_0) \right]^\top \gamma$.

\section{Proof of Proposition \ref{prop:asymptotic}} \label{sec:proof:asymptotic}

Recall that we obtain the estimator $(\hat \beta, \hat \gamma)$ by solving $\PP_n U(\beta, \gamma) = 0$, where
\begin{align*}
\begin{split} 
& U (\beta, \gamma) = \sum\limits_{\widebar d \in \cD} \sum\limits_{t=1}^T \left\{ \vphantom{\sum\limits_{\widebar d \in \cD} \sum\limits_{t=1}^T} W_{\widebar d}^S W_t^M
\left( Y_{t+1} - (g_t(H_t) - \mu_{t,\barz} (H_t))^\top \alpha \right. \right. \\
& - \left.\left.(A_t - \tilde{p}_t(1 \mid H_t)) f_t \left( \widebar{d}, x_0 \right)^\top \beta - m_t \left( \widebar{d}, x_0 \right)^\top \gamma \right) 
\begin{pmatrix}
(g_t(H_t) - \mu_{t,\barZ, R} (H_t)) \\
(A_t - \tilde{p}_t(A_t \mid H_t)) f_t \left( \widebar{d}, x_0 \right) \\
m_t (\bar{Z}_t)
\end{pmatrix}\right\}
\end{split},
\end{align*}
where $\mu_{\PP_n,t} (H_t) = \frac{\PP_n \sum\limits_{\widebar d \in \cD} W_{\widebar d}^S g_t(H_t)}{\PP_n \sum\limits_{\widebar d \in \cD} W_{\widebar d}^S}$.

We make the following regularity and moments conditions.
\begin{assumption}
\begin{enumerate}[label = (\arabic*)]
\item \hspace{1em} All entries in $\{Y_{t+1}, g_t(H_t)\}_{t=1}^{T}$ have finite fourth moments.

\item \hspace{1em} Define $\mu_t (H_t) = \frac{\EE \left[ \sum\limits_{\widebar d \in \cD} W_{\widebar d}^S g_t(H_t) \right]}{\EE \left[ \sum\limits_{\widebar d \in \cD} W_{\widebar d}^S \right]} = \frac{1}{D} \EE \left[ \sum\limits_{\widebar d \in \cD} W_{\widebar d}^S g_t(H_t) \right]$, where $D = |\cD|$ is the number of DTRs embedded in he SMART design. Also denote $U_2 (\mu_t) = \sum\limits_{\widebar d \in \cD} W_{\widebar d}^S (\mu_{t} (H_t) - g_t(H_t))$.
The centering control variable satisfies $\mu_{\PP_n,t} (H_t) = \mu_t (H_t) + o_p(1)$, and $\sqrt{n} \left(\mu_{\PP_n,t} (H_t) - \mu_t (H_t) \right) = D^{-1} \PP_n U_2 (\mu_t) + o_p(1)$. Therefore, $\sqrt{n} \left(\mu_{\PP_n,t} (H_t) - \mu_t (H_t) \right)$ converges in distribution to a normal distribution with mean 0 and variance $D^{-2} \EE \left[ U_2 ^{\otimes 2} \right]$, which has finite entries.
\end{enumerate}
\end{assumption}

Under standard regularity and moments conditions, we have the solution $(\hat \alpha, \hat{\beta}, \hat \gamma)^\top - (\alpha, \beta, \gamma)^\top$ converge asymptotically in distribution to a Gaussian distribution with mean 0 as $n \rightarrow \infty$, where $(\alpha, \beta, \gamma)$ is the solution to equation $\EE U(\beta, \gamma) = 0$. The variance of the asymptotic Gaussian distribution is given by 
$$\left[ \EE \dot U (\beta, \gamma) \right]^{-1} \left[ \EE U (\beta, \gamma)^{\otimes 2} \right] \left[ \EE \dot U (\beta, \gamma) \right]^{-1},$$
where $\EE \left[ \dot U (\beta, \gamma) \right] = $
\[
\EE\left[ \begin{pmatrix}
\sum\limits_{\widebar d \in \cD} \sum\limits_{t=1}^T W_{\widebar d}^S \tilde{p}_t(1 \mid H_t) (1 - \tilde{p}_t(1 \mid H_t)) f_t \left( \widebar{d}, x_0 \right)^{\otimes 2} & 0\\
0 & \sum\limits_{\widebar d \in \cD} \sum\limits_{t=1}^T W_{\widebar d}^S m_t \left( \widebar{d}, x_0 \right)^{\otimes 2} 
\end{pmatrix}\right],
\]
\begin{align*}
\zeta_t &= \left( Y_{t+1} - (g_t(H_t) - \EE[g_t(H_t) \mid \bar{Z}])^\top \alpha - (A_t - \tilde{p}_t(1 \mid H_t)) f_t \left( \widebar{d}, x_0 \right)^\top \beta - m_t \left( \widebar{d}, x_0 \right)^\top \gamma \right), \\
\EE \left[ U (\beta, \gamma)^{\otimes 2} \right] &= \EE \left[ \sum\limits_{\widebar d \in \cD} \sum\limits_{t=1}^T W_{\widebar d}^S W_t^M \zeta_t
\begin{pmatrix}
(A_t - \tilde{p}_t(A_t \mid H_t)) f_t \left( \widebar{d}, x_0 \right) \\
m_t (\bar{Z}_t)
\end{pmatrix}
\right]^{\otimes 2}.
\end{align*}
A consistent estimator of the asymptotic variance is given by 
$\left[ \PP_n \dot U (\hat{\beta}, \hat{\gamma}) \right]^{-1}
\left[ \PP_n U (\hat{\beta}, \hat{\gamma}) \right]^{\otimes 2}
\left[ \PP_n \dot U (\hat{\beta}, \hat{\gamma}) \right]^{-1}$.

Next, we show that the limit $(\beta, \gamma)$ equals the true causal coefficient $(\beta^*, \gamma^*)$. 
For simplicity, we abbreviate the centered control variable $g_t(H_t) - \EE[g_t(H_t) \mid \bar{Z}]$ as $g^c_t(H_t)$. We first solve for $\alpha$ that satisfies
\begin{align*}
& \EE \left[ \left\{ \sum\limits_{\widebar d \in \cD} \sum\limits_{t=1}^T W_{\widebar d}^S W_t^M
\left( Y_{t+1} - g^c_t(H_t)^\top \alpha \right. \right. \right. \\
& - \left.\left.\left.(A_t - \tilde{p}_t(1 \mid H_t)) f_t \left( \widebar{d}, x_0 \right)^\top \beta - m_t \left( \widebar{d}, x_0 \right)^\top \gamma \right) 
\vphantom{\sum\limits_{\widebar d \in \cD} \sum\limits_{t=1}^T}
\right\} g^c_t(H_t) \right] = 0.
\end{align*}
Note that 
\[
\EE \left[ \sum\limits_{\widebar d \in \cD} \sum\limits_{t=1}^T W_{\widebar d}^S W_t^M (A_t - \tilde{p}_t(1 \mid H_t)) f_t \left( \widebar{d}, x_0 \right)^\top \beta g^c_t(H_t) \right] = 0,
\]
\[ 
\EE \left[ \sum\limits_{\widebar d \in \cD} \sum\limits_{t=1}^T W_{\widebar d}^S W_t^M m_t \left( \widebar{d}, x_0 \right)^\top \gamma g^c_t(H_t) \right] = 0.
\]
Hence
\begin{align*}
\alpha = \EE \left[ \sum\limits_{\widebar d \in \cD} \sum\limits_{t=1}^T W_{\widebar d}^S g^c_t(H_t)^{\otimes 2} \right]^{-1}
\EE \left[ \sum\limits_{\widebar d \in \cD} \sum\limits_{t=1}^T W_{\widebar d}^S g^c_t(H_t) \EE \left[ W_t^M Y_{t+1} \mid H_t \right] \right].
\end{align*}

We next solve for $\beta$ that satisfies 
\begin{align*}
& \EE \left[ \left\{ \sum\limits_{\widebar d \in \cD} \sum\limits_{t=1}^T W_{\widebar d}^S W_t^M
\left( Y_{t+1} - g^c_t(H_t)^\top \alpha \right. \right. \right. \\
& - \left.\left.\left.(A_t - \tilde{p}_t(1 \mid H_t)) f_t \left( \widebar{d}, x_0 \right)^\top \beta - m_t \left( \widebar{d}, x_0 \right)^\top \gamma \right) 
\vphantom{\sum\limits_{\widebar d \in \cD} \sum\limits_{t=1}^T}
\right\} (A_t - \tilde{p}_t(1 \mid H_t)) f_t \left( \widebar{d}, x_0 \right) \right] = 0.
\end{align*}
Note that 
\[
\EE \left[ \sum\limits_{\widebar d \in \cD} \sum\limits_{t=1}^T W_{\widebar d}^S W_t^M g^c_t(H_t)^\top \alpha (A_t - \tilde{p}_t(1 \mid H_t)) f_t \left( \widebar{d}, x_0 \right) \right] = 0,
\]
\[ 
\EE \left[ \sum\limits_{\widebar d \in \cD} \sum\limits_{t=1}^T W_{\widebar d}^S W_t^M m_t \left( \widebar{d}, x_0 \right)^\top \gamma (A_t - \tilde{p}_t(1 \mid H_t)) f_t \left( \widebar{d}, x_0 \right) \right] = 0,
\]
\begin{align}
& \EE \left[ W_{\widebar d}^S W_t^M Y_{t+1} (A_t - \tilde{p}_t(1 \mid H_t)) f_t \left( \widebar{d}, x_0 \right) \right] \nonumber \\
=& W_{\widebar d}^S \tilde{p}_t(1 \mid H_t) (1 - \tilde{p}_t(1 \mid H_t))
\underbrace{\left( \EE \left[ Y_{t+1} \mid \barZ = \barz, A_t = 1 \right] - \EE \left[ Y_{t+1} \mid \barZ = \barz, A_t = 0 \right] \right)}_{\delta_A^{\bar z}}, \label{eq::derivation:EY} \\
& \EE \left[ W_{\widebar d}^S W_t^M (A_t - \tilde{p}_t(1 \mid H_t))^2 f_t \left( \widebar{d}, x_0 \right)^\top \beta f_t \left( \widebar{d}, x_0 \right) \right] \nonumber \\
=& W_{\widebar d}^S \tilde{p}_t(1 \mid H_t) (1 - \tilde{p}_t(1 \mid H_t))
 f_t \left( \widebar{d}, x_0 \right)^\top \beta f_t \left( \widebar{d}, x_0 \right). \nonumber 
\end{align}
Therefore, we have 
\begin{align}
\begin{split} \label{eq:beta:form}
\beta = &\EE \left[ \sum\limits_{\widebar d \in \cD} \sum\limits_{t=1}^T W_{\widebar d}^S \tilde{p}_t(1 \mid H_t) (1 - \tilde{p}_t(1 \mid H_t)) f_t \left( \widebar{d}, x_0 \right)^{\otimes 2} \right]^{-1} \\
&\hspace{2em} \EE \left[ \sum\limits_{\widebar d \in \cD} \sum\limits_{t=1}^T W_{\widebar d}^S \tilde{p}_t(1 \mid H_t) (1 - \tilde{p}_t(1 \mid H_t)) \delta_A^{\bar z} f_t \left( \widebar{d}, x_0 \right) \right].
\end{split}
\end{align}

Under Assumption \ref{assumption:identification}, we have $\EE \left[ Y_{t+1} \left( \barz, (\bar{A}_{t-1}, 1) \right) - Y_{t+1} \left( \barz, (\bar{A}_{t-1}, 0) \right) \right] 
= \EE \left[ Y_{t+1} \mid \barZ = \barz, A_t = 1 \right] - \EE \left[ Y_{t+1} \mid \barZ = \barz, A_t = 0 \right] = \delta_A^{\bar z}$. Therefore,
If the model \eqref{eq:marginal:model:eq1} is correct, we have 
\begin{align*} 
\beta &= \EE \left[ \sum\limits_{\widebar d \in \cD} \sum\limits_{t=1}^T W_{\widebar d}^S \tilde{p}_t(1 \mid H_t) (1 - \tilde{p}_t(1 \mid H_t)) f_t \left( \widebar{d}, x_0 \right)^{\otimes 2} \right]^{-1} \\
&\ \EE \left[ \sum\limits_{\widebar d \in \cD} \sum\limits_{t=1}^T W_{\widebar d}^S \tilde{p}_t(1 \mid H_t) (1 - \tilde{p}_t(1 \mid H_t)) f_t \left( \widebar{d}, x_0 \right) f_t \left( \widebar{d}, x_0 \right)^\top \beta^* \right] = \beta^*.
\end{align*}

Similarly we can solve for $\gamma$ that satisfies 
\begin{align*}
& \EE \left[ \left\{ \sum\limits_{\widebar d \in \cD} \sum\limits_{t=1}^T W_{\widebar d}^S W_t^M
\left( Y_{t+1} - g^c_t(H_t)^\top \alpha \right. \right. \right. \\
& - \left.\left.\left.(A_t - \tilde{p}_t(1 \mid H_t)) f_t \left( \widebar{d}, x_0 \right)^\top \beta - m_t \left( \widebar{d}, x_0 \right)^\top \gamma \right) 
\vphantom{\sum\limits_{\widebar d \in \cD} \sum\limits_{t=1}^T}
\right\} m_t(\barZ) \right] = 0,
\end{align*}
which results in
\begin{align*}
\gamma = \EE \left[ \sum\limits_{\widebar d \in \cD} \sum\limits_{t=1}^T W_{\widebar d}^S m_t(\barZ)^{\otimes 2} \right]^{-1}
\EE \left[ \sum\limits_{\widebar d \in \cD} \sum\limits_{t=1}^T W_{\widebar d}^S m_t(\barZ) \EE \left[ W_t^M Y_{t+1} \mid H_t \right] \right].
\end{align*}

\section{Comparison between Causal Estimands Accounting for Availability} \label{sec:method:availability:compareWCLS}

Let $S_t := I_t$ and the centered availability be $I_t^c = I_t - \psi_t$, where $\psi_t = \frac{\sum_{\bar d \in \cD} \sum_{t=1}^T 
W_{\widebar d}^S
\tilde p_t(1) (1 - \tilde p_t(1)) I_t}
{\sum_{\bar d \in \cD} \sum_{t=1}^T W_{\widebar d}^S \tilde p_t(1) (1 - \tilde p_t(1))}$. 
We consider the form of $\alpha_1$. Recall that 
\begin{align}
\begin{split} \label{eq:alpha1:derive1} 
0 =& \EE \left[
\sum_{\bar d \in \cD} \sum\limits_{t=1}^T 
W_{\widebar d}^S W_t^M
\left[ Y_{t+1} - g_t^c(H_t)^\top \alpha_0 \right.\right. \\
&\hspace{4em} - \left.\left. (A_t - \rho) \left( f_t (\widebar{d}, x_0)^\top \beta + I_t^c \alpha_1 \right) 
- m_t (\widebar{d}, s)^\top \eta 
\vphantom{\left( g_t(H_t) - \mu_{t,\widebar d} (H_t) \right)^\top} \right] 
(A_t - \rho) I_t^c \right].
\end{split}
\end{align}
Then \eqref{eq:alpha1:derive1} equals 
\begin{align*}
0 &= \EE \left\{
\sum_{\bar d \in \cD} \sum\limits_{t=1}^T 
W_{\widebar d}^S W_t^M
\left[ Y_{t+1} - g_t^c(H_t)^\top \alpha_0 \right.\right. \\
&\hspace{4em} - \left.\left. (A_t - \rho) \left( f_t (\widebar{d}, x_0)^\top \beta + I_t^c \alpha_1 \right) 
- m_t (\widebar{d}, s)^\top \eta 
\right] 
\vphantom{\sum_{\bar d \in \cD} \sum\limits_{t=1}^T } (A_t - \rho) I_t^c \right\} \\
&= \EE \left\{
\sum_{\bar d \in \cD} \sum\limits_{t=1}^T 
W_{\widebar d}^S W_t^M
\left[ \EE [Y_{t+1} \mid H_t, A_t, I_t] - g_t^c(H_t)^\top \alpha_0 \right.\right. \\
&\hspace{4em} - \left.\left. (A_t - \rho) \left( f_t (\widebar{d}, x_0)^\top \beta + I_t^c \alpha_1 \right) 
- m_t (\widebar{d}, s)^\top \eta 
\right] 
\vphantom{\sum_{\bar d \in \cD} \sum\limits_{t=1}^T } 
(A_t - \rho) I_t^c \right\}  \\
&= \EE \left\{
\sum_{\bar d \in \cD} \sum\limits_{t=1}^T 
W_{\widebar d}^S W_t^M
\left[ \vphantom{f_t (\widebar{d}, x_0)^\top} 
\EE [Y_{t+1} \mid H_t, A_t, I_t] \right.\right. \\
&\hspace{4em} - \left.\left. (A_t - \rho) \left( f_t (\widebar{d}, x_0)^\top \beta + I_t^c \alpha_1 \right) 
\right] 
\vphantom{\sum_{\bar d \in \cD} \sum\limits_{t=1}^T } 
(A_t - \rho) I_t^c \right\}  \\
&= \EE \left\{
\sum_{\bar d \in \cD} \sum\limits_{t=1}^T 
W_{\widebar d}^S
\sum_{i=0}^1 \sum_{a=0}^1 
\left[ \vphantom{f_t (\widebar{d}, x_0)^\top} 
W_t^M \EE [Y_{t+1} \mid H_t, A_t = a, I_t = i] - (a - \rho) \left( f_t (\widebar{d}, x_0)^\top \beta + (i - \psi_t) \alpha_1 \right) 
\right] \right. \\
&\hspace{4em} \left. \vphantom{\sum_{\bar d \in \cD} \sum\limits_{t=1}^T } 
(a - \rho) (i - \psi_t) P(A_t = a \mid H_t, I_t = i) P(I_t = i \mid H_t) \right\},
\end{align*}
where the last equality averages over $A_t$ and $I_t$.
When $I_t = 0$, the MRT treatment $A_t$ is not randomized and we let $A_t = \tilde{p}_t(1) = \rho$ in this case. Therefore $P(A_t = \rho \mid H_t, I_t = 0) = 1$ and $P(A_t = a \mid H_t, I_t = 0) = 0$ for $a \in \{0, 1\}$. 
The last equality then simplifies to 
\begin{align*}
0 &= 
%
\EE \left\{
\sum_{\bar d \in \cD} \sum\limits_{t=1}^T 
W_{\widebar d}^S
\left[ \vphantom{f_t (\widebar{d}, x_0)^\top} 
 (\EE [Y_{t+1} \mid H_t, A_t = 1, I_t = 1] -  \EE [Y_{t+1} \mid H_t, A_t = 0, I_t = 1])
\right.\right. \\
&\hspace{4em} - \left.\left. \left( f_t (\widebar{d}, x_0)^\top \beta + (1 - \psi_t) \alpha_1 \right) 
\right] 
\vphantom{\sum_{\bar d \in \cD} \sum\limits_{t=1}^T } 
\tilde{p}_t(1) (1 - \tilde{p}_t(1))
(1 - \psi_t) P(I_t = 1 \mid H_t) \right\} \\
&= \EE \left\{
\sum_{\bar d \in \cD} \sum\limits_{t=1}^T 
W_{\widebar d}^S
\left[ \vphantom{f_t (\widebar{d}, x_0)^\top} 
 (\EE [ \EE [Y_{t+1} \mid H_t, A_t = 1, I_t = 1] -  \EE [Y_{t+1} \mid H_t, A_t = 0, I_t = 1] \mid \widebar Z, I_t = 1])
\right.\right. \\
&\hspace{4em} - \left.\left. \left( f_t (\widebar{d}, x_0)^\top \beta + (1 - \psi_t) \alpha_1 \right) 
\right] 
\vphantom{\sum_{\bar d \in \cD} \sum\limits_{t=1}^T } 
\tilde{p}_t(1) (1 - \tilde{p}_t(1))
(1 - \psi_t) P(I_t = 1 \mid H_t) \right\}
\end{align*}

Thus
\begin{align} 
\begin{split} \label{eq:alpha1:availability}
\alpha_1 &= \EE \left[ \sum\limits_{\widebar d \in \cD} \sum\limits_{t=1}^T W_{\widebar d}^S \tilde{p}_t(1) (1 - \tilde{p}_t(1)) (1 - \psi_t)^2 P(I_t = 1 \mid H_t) \right]^{-1} \\
&\hspace{4em}  \EE \left[ \sum\limits_{\widebar d \in \cD} \sum\limits_{t=1}^T W_{\widebar d}^S 
\left( \delta_A^{\bar Z, I=1} - \delta_A^{\bar Z} \right) 
(1 - \psi_t) P(I_t = 1 \mid H_t) \right],    
\end{split}
\end{align}
where 
\begin{align*}
\delta_A^{\widebar Z, I = 1} &= \EE [ \EE [Y_{t+1} \mid H_t, A_t = 1, I_t = 1] -  \EE [Y_{t+1} \mid H_t, A_t = 0, I_t = 1] \mid \widebar Z, I_t = 1], \\
\delta_A^{\widebar Z} &= \EE \left[ \EE \left[ Y_{t+1} \mid H_t, A_t = 1 \right] - \EE \left[ Y_{t+1} \mid H_t, A_t = 0 \right] \middle\vert \widebar Z \right].
\end{align*}
The contrast $\delta_A^{\widebar Z, I = 1} - \delta_A^{\widebar Z}$ represents the causal effect of the MRT treatment for a fixed DTR attributed to accounting for availability status.
If availability $I_t$ is conditionally mean independent of the MRT treatment effect given $\widebar Z$, then $\delta_A^{\widebar Z, I = 1} = \delta_A^{\widebar Z}$ and $\alpha_1 = 0$. 
In the M-Bridge study, however, availability is predictive of the SMART treatment in the second stage. This is because if $I_t = 0$ for any $t \in \{2, 3, 4\}$ when a student is classified as a heavy drinker, then $Z_2$ must be a resource email $(1)$ or an online health coach $(-1)$, while $I_t = 1$ for all $t \in \{1,2,3,4\}$ implies $Z_2 = 0$. Therefore, based on Proposition 4.4 of \cite{shi2023IncorporatingAuxiliary}, accounting for availability status improves efficiency when estimating the causal effect.

On the other hand, \eqref{eq:alpha1:availability} suggests that the proposed estimator for the hybrid design accounts for availability in a different manner than WCLS \citep{boruvka2018assessing}. 
Recall that WCLS estimates the causal excursion effect of MRT treatments by conditioning on available time points and assumes the model 
$$ \EE [ \EE [Y_{t+1} \mid H_t, A_t = 1, I_t = 1] -  \EE [Y_{t+1} \mid H_t, A_t = 0, I_t = 1] \mid I_t = 1] = f(X_0)^\top \beta^{WCLS}. $$
Using the WCLS method, $\beta^{WCLS} =$
\begin{align*}
&\EE \left[ \sum\limits_{t=1}^T \tilde{p}_t(1) (1 - \tilde{p}_t(1)) f_t \left( x_0 \right)^{\otimes 2} I_t \right]^{-1} 
\EE \left[ \sum\limits_{t=1}^T W_{\widebar d}^S \tilde{p}_t(1) (1 - \tilde{p}_t(1)) \right. \\
&\hspace{2em}  \left. \EE \left[ \EE [Y_{t+1} \mid H_t, A_t = 1, I_t = 1] -  \EE [Y_{t+1} \mid H_t, A_t = 0, I_t = 1] \mid I_t = 1 \right] f_t \left( x_0 \right) I_t
\vphantom{\sum_{\bar d \in \cD} \sum\limits_{t=1}^T } 
\right].
\end{align*}

The corresponding parameter in the hybrid method takes the form as in \eqref{eq:beta:form}:
\begin{align*}
\beta = &\EE \left[ \sum\limits_{\widebar d \in \cD} \sum\limits_{t=1}^T W_{\widebar d}^S \tilde{p}_t(1) (1 - \tilde{p}_t(1)) f_t \left( \widebar{d}, x_0 \right)^{\otimes 2} \right]^{-1} \\
&\hspace{2em} \EE \left[ \sum\limits_{\widebar d \in \cD} \sum\limits_{t=1}^T W_{\widebar d}^S \tilde{p}_t(1) (1 - \tilde{p}_t(1)) \delta_A^{\widebar Z} f_t \left( \widebar{d}, x_0 \right) \right].
\end{align*}

Therefore, $\beta^{WCLS}$ only uses observations at time points when $A_t$ is delivered and yields a conditional effect given availability. In contrast, $\beta$ uses observations at all time points and yields a marginal effect averaging out availability which is accounted for as an auxiliary variable via \eqref{eq:alpha1:availability}.

\clearpage 
\newpage
\section{Scientific Questions for A Hybrid Design} \label{sec:questions}

We list additional scientific questions of interest to a hybrid SMART-MRT study in Table \ref{tab:questions}.

\begin{table}[H] 
\caption{Scientific questions about proximal effects and model parameters for a hybrid SMART-MRT design} \label{tab:questions}
\footnotesize
\rotatebox{-90}{
\begin{tabularx}{1.4\linewidth}{lXlXX}
\toprule
& Scientific question & Type & Estimand & 
\makecell{Model parameters \\ in Example \ref{ex:marginal:model}} \\
\midrule
(a) & Does the effects of two DTRs, e.g., $\widebar d = (1, 1)$ vs $(1, -1)$, on the proximal outcome at time $t \geq t^*$ differ for a fixed MRT treatment $A_t = 1$? & Interaction &
$\!\begin{aligned}[t]
\EE & \left[ Y_{t+1} \left( \widebar d = (1, 1), (\widebar{A}_{t-1}, 1) \right) \right. \\
 & - \left. Y_{t+1} \left( \widebar d = (1, -1), (\widebar{A}_{t-1}, 1) \right) \right]
\end{aligned}$
 & $\!\begin{aligned}[t]
( & \beta_0 + \beta_1 + 2 \beta_2 + 2 \beta_3) / 2 \\
& + 2 \eta_2 + 2 \eta_3 
\end{aligned}$ \\
\hline
(b) & Does the effect of MRT treatment on the proximal outcome at time $t < t^*$ vary by first-stage regimes, $d_1 = 1$ vs $d_1 = -1$? & Interaction &
$\!\begin{aligned}[t]
\EE & \left[ Y_{t+1} \left( d_1 = 1, (\widebar{A}_{t-1}, 1) \right) \right. \\
 & - \left. Y_{t+1} \left( d_1 = 1, (\widebar{A}_{t-1}, 0) \right) \right] \\
-\EE & \left[ Y_{t+1} \left( d_1 = -1, (\widebar{A}_{t-1}, 1) \right) \right. \\
 & - \left. Y_{t+1} \left( d_1 = -1, (\widebar{A}_{t-1}, 0) \right)  \right] 
\end{aligned}$
 & $2 \beta_1$ \\
\hline
(c) & Does the effect of MRT treatment on the proximal outcome at time $t > t^*$ vary by DTRs, e.g., $\widebar d = (1, 1)$ vs $(1, -1)$? & Interaction &
$\!\begin{aligned}[t]
\EE & \left[ Y_{t+1} \left( \widebar d = (1, 1), (\widebar{A}_{t-1}, 1) \right) \right. \\
& - \left. Y_{t+1} \left( \widebar d = (1, 1), (\widebar{A}_{t-1}, 0) \right) \right] \\
-\EE & \left[ Y_{t+1} \left( \widebar d = (1, -1), (\widebar{A}_{t-1}, 1) \right) \right. \\
& - \left. Y_{t+1} \left( \widebar d = (1, -1), (\widebar{A}_{t-1}, 0) \right)  \right] 
\end{aligned}$
 & $2 \beta_2 + 2 \beta_3$ \\
\hline
(d) & Do the effects of MRT treatments differ on the proximal outcome at time $t > t^*$ averaging over DTRs? & Main &
$\!\begin{aligned}[t]
\sum_{\bar d \in \cD} P(\bar d)  \left\{ 
\EE \right. &\left[ Y_{t+1} \left( \widebar d, (\widebar{A}_{t-1}, 1) \right) \right. \\
& \left. \left. - Y_{t+1} \left( \widebar d, (\widebar A_{t-1}, 0) \right) \right] \right\}
\end{aligned}$ 
& $\beta_0 $ \\
\hline
(e) & Do the effects of two DTR, e.g., $\widebar d = (1, 1)$ and $(1, 1)$, differ on the proximal outcome at time $t > t^*$ averaging over MRT treatments? & Main &
$\!\begin{aligned}[t]
\EE & \left[ Y_{t+1} \left( \widebar d = (1, 1), \widebar{A}_{t} \right) \right. \\
& - \left. Y_{t+1} \left( \widebar d = (1, -1), \widebar{A}_{t} \right) \right]
\end{aligned}$ 
& $2\gamma_2 + 2 \gamma_3 $ \\
\hline
\bottomrule
\end{tabularx}}
\end{table}

\newpage
\section{Additional Details of Simulations} \label{sec:sim:additional}

\subsection{Detailed Simulation Scenarios} \label{sec:sim:setup:additional}
In scenario II, we let $p_t (1 \mid H_t) = p_t (1 \mid Z_1, R, Z_2) = 0.6 I\{ Z_1 = 1 \} + 0.4 I\{ Z_1 = -1 \} - 0.2 I\{ C_t Z_2 = 1 \} + 0.2 I\{ C_t Z_2 = -1 \}$ be DTR-dependent, 
The responder status is history-dependent, with 
$\mathrm{logit} (p(R = 1 \mid H_{t^*})) = -0.62 + \tilde X_1 + (A_{t^*-1} - p_{t^*-1}(1 \mid H_{t^*-1})) + 0.5 Z_1$, where 
$\tilde{X}_t = X_{t} - \EE [X_{t} \mid A_{t-1}, H_{t-1}]$ is the centered state. The realized values of the probability of being a responder are about 0.45 and 0.27 for individuals with $Z_1 = 1$ and $-1$, respectively.

\subsection{Baseline Model Setup in Simulations} \label{sec:sim:baseline:model}
In WCLS, the numerator probability in the MRT weight was set as $\tilde p_t(1) = \frac{1}{2}$. The control variables $g_{t} (H_t)$ included non-centered $X_t$ and $X_t Z_1$. 
The marginal model of WCLS analysis included DTRs as moderators is specified as 
\begin{align}
  \begin{split} \label{eq:sim:wcls:eq}
    & E[ Y_{t+1} \left( \widebar d, (\bar A_{t-1}, 1) \right) - Y_{t+1} \left( \widebar d, (\bar A_{t-1}, 0) \right)] = f_t (\widebar d)^\top \beta_{\mathrm{WCLS}}.
\end{split} 
\end{align}
The coefficient $\hat{\beta}_{\mathrm{WCLS}}$ is obtained by solving the estimating equation
\begin{align}
  \begin{split} \label{eq:sim:wcls:ee}
  & \PP_n \sum\limits_{t=1}^T W_t^M
  \left[ Y_{t+1} - g_t(H_t)^\top \alpha_0 - (A_t - \rho) f_t (\widebar d)^\top \beta_{\mathrm{WCLS}} \right]
  \begin{pmatrix}
  g_t(H_t) - \mu_{t,\widebar d} (H_t) \\
  (A_t - \rho) f_t \left( \widebar{d}\right)
  \end{pmatrix} = 0.
\end{split}
\end{align}
Note that in \eqref{eq:sim:wcls:ee}, the control variables $g_t(H_t)$ are not centered as in \eqref{eq:ee:beta} of the proposed approach.
In both simulation scenarios I and II, we let $f_t (\widebar d) = (1, d_1, \delta_t d_2, \delta_t d_1 d_2)^\top$, taking the same form as in the hybrid method.

The WR method focuses only on the marginal mean proximal outcome under a DTR averaging over all MRT treatment sequences, and does not account for time-varying control variables. The marginal mean model is specified as 
\begin{align}
\begin{split} \label{eq:sim:wr:model}
    & E[ Y_{t+1} \left( \widebar d, \bar A_{t} \right)] = m_t (\widebar d)^\top \beta_{\mathrm{WR}}.
\end{split} 
\end{align}
The coefficient $\hat{\beta}_{\mathrm{WR}}$ is obtained by solving the estimating equation
\begin{align}
  \begin{split} \label{eq:sim:wr:ee}
  & \PP_n \sum\limits_{\widebar d \in \cD} \sum\limits_{t=1}^T W_t^S
  \left[ Y_{t+1} - m_t (\widebar d)^\top \beta_{\mathrm{WR}} \right] m_t \left( \widebar{d}\right)
    = 0.
\end{split}
\end{align}
In scenario I, $m_t (\widebar d) = (1, d_1, \delta_t d_2, \delta_t d_1 d_2)^\top$; in scenario II, $m_t (\widebar d) = (1-\delta_t, (1-\delta_t) d_1, \delta_t, \delta_t d_1, \delta_t d_2, \delta_t d_1 d_2)^\top$.
The moderators $f_t$ were the same as \eqref{eq:example:model:interaction} in both scenarios I and II, respectively. 


\subsection{Expression of True Marginal Effects in Simulation} \label{sec:sim:expression}

We consider three types of marginal quantities: (1) the marginal means of the proximal outcome at a fixed DTR and $A_t = a$, (2) the marginal means of the proximal outcome for a given $A_t = a$, averaging over all DTRs, and (3) the marginal means of the proximal outcome for a given DTR, averaging over all $A_t$'s.

\subsubsection{Marginal Mean At A Fixed DTR And $A_t = a$}
We compute $\EE[Y(A_t = a, \bar Z = \bar z]$ for all $a \in \{0, 1\}$ and $\bar z \in \cD$. Based on the causal assumptions, we know that $\EE[Y_{t+1} ((\bar A_{t-1}, a), \bar Z = \bar z)] = \EE[Y_{t+1} \mid A_t = a, \bar Z = \bar z]$. In addition, since $\EE [\tilde X_t] = 0$ for all $1 \leq t \leq T$ in the data generating process \eqref{eq:sim:dgp}, we have that $\EE [(A_{t} - p_{t}(1 \mid H_{t})) \tilde X_t] = 0$, $\EE [(A_{t} - p_{t}(1 \mid H_{t})) \tilde X_t Z_1] = 0$, and $\EE [\tilde X_t Z_1] = 0$.
Recall that $H_t^M = H_t \setminus \{\bar Z\}$ denotes the history up to time $t$ except for SMART factors.

We now consider Stage 1 and Stage 2 separately. 

\paragraph*{Stage 1}
When $t < t^*$, 
\begin{align}
& \EE[ Y_{t+1} \left((\bar A_{t-1}, a), D_1 = d_1 \right)] \nonumber \\
=& \sum_{z_1 \in \{-1, 1\}}  I\{z_1 = d_1 (X_0)\} \EE[ Y_{t+1} \left((\bar A_{t-1}, a), Z_1 = z_1 \right)] \nonumber \\
=& \sum_{z_1 \in \{-1, 1\}}  I\{z_1 = d_1 (X_0)\} \EE[ \EE [Y \mid A_t = a, Z_1 = z_1, H_t^M] ] \nonumber \\
=& \sum_{z_1 \in \{-1, 1\}}  I\{z_1 = d_1 (X_0)\} \left[ (a - p_{t}(1 \mid z_1)) (\beta_0^* + \beta_1^* z_{1}) + \gamma_0^* + \gamma_1^* z_{1} \right]. \label{eq:sim:true:fixAZ:stage1}
\end{align}

\paragraph*{Stage 2}
When $t > t^*$, 
\begin{align}
& \EE \left[ Y_{t+1} \left(\bar d = (d_1, d_2), (\bar A_{t-1}, a) \right) \right]  \nonumber \\
=& \sum_{z \in \cZ} \EE \left[ I\{z_1 = d_1 (X_0)\} I\{z_2 = d_2 (H_{t^*} (z_1))\} 
Y \left((\bar A_{t-1}, a), \bar Z = (z_1, z_2) \right) \right] \nonumber \\
=& \sum_{z \in \cZ} \sum_{r=0}^1 \EE \left[ I\{z_1 = d_1 (X_0)\} I\{z_2 = d_2 (H_{t^*}(z_1), R(z_1) = r)\} \right. \nonumber \\
&\ \ \ \left.  Y_{t+1} \left( (\bar A_{t-1}, a), \bar Z = (z_1, z_2), R(z_1) = r \right) P(R(z_1) = r) \right] \nonumber \\
=& \sum_{z \in \cZ} \sum_{r=0}^1 \EE \left[ I\{z_1 = d_1 (X_0)\} I\{z_2 = d_2 (Z_1 = z_1, R = r)\} \right.\nonumber \\
&\ \ \ \left. \EE \left[ Y_{t+1} \mid A_t = a, \bar Z = (z_1, z_2), R = r, H_t^M \right] P(R = r \mid Z_1 = z_1) \right] \nonumber \\
\begin{split}\label{eq:sim:true:fixAZ:stage2}
=& \sum_{z \in \cZ} I\{z_1 = d_1 (X_0)\} I\{z_2 = d_2 (Z_1 = z_1, R = r)\} \\
&\ \ \left\{ \left[ (A_{t} - p_{t}(1 \mid \bar Z = (z_1, z_2), R = 0)) (\beta_0^* + \beta_1^* z_{1} + \beta_2^* z_{2} + \beta_3^* z_{1} z_{2}) +
\gamma_0^* + \gamma_1^* z_{1} + \gamma_2^* z_{2}\right. \right. \\
&\ \ \left. + \gamma_3^* z_{1} z_{2} \right] \cdot P(R = 0 \mid Z_1 = z_1) + \\
&\ \ \left. \left[ (A_{t} - p_{t}(1 \mid \bar Z = (z_1, 0), R = 1)) (\beta_0^* + \beta_1^* z_{1}) +
\gamma_0^* + \gamma_1^* z_{1} \right] \cdot P(R = 1 \mid Z_1 = z_1) \right\}.
\end{split}
\end{align}


\subsubsection{Marginal Mean at A Fixed $A_t = a$ Averaging over DTRs}

\paragraph*{Stage 1}
When $t < t^*$, since we have equal probabilities of assigning two SMART factors, the marginal means averaging over all DTRs can be expressed as
\begin{align*}
& \sum_{d_1 \{-1, 1\}} \EE[ Y_{t+1} \left((\bar A_{t-1}, a), d_1 \right)] P(D_1 = d_1) \\
=& \frac{1}{2} \EE[ Y_{t+1} \left((\bar A_{t-1}, a), 1 \right)] + \frac{1}{2} \EE[ Y_{t+1} \left((\bar A_{t-1}, a), -1 \right)] \\
=& \frac{1}{2} \left[ (a - p_{t}(1 \mid z_1 = 1)) (\beta_0^* + \beta_1^*) + \gamma_0^* + \gamma_1^* \right] + \\ 
&\ \ \frac{1}{2} \left[ (a - p_{t}(1 \mid z_1 = -1)) (\beta_0^* - \beta_1^*) + \gamma_0^* - \gamma_1^* \right] \\
=& \frac{1}{2} \left[ (a - p_{t}(1 \mid z_1 = 1)) (\beta_0^* + \beta_1^*) + (a - p_{t}(1 \mid z_1 = -1)) (\beta_0^* - \beta_1^*) \right] + \gamma_0^*.
\end{align*}
As a result, the marginal effect of the MRT treatment on the proximal outcome averaging over DTRs is
$\sum_{d_1 \{-1, 1\}} \EE[ Y_{t+1} \left((\bar A_{t-1}, 1), d_1 \right) - Y_{t+1} \left((\bar A_{t-1}, 0), d_1 \right)] P(D_1 = d_1) = \beta_0^*$.

\paragraph*{Stage 2}
When $t > t^*$, the marginal means averaging over all DTRs can be expressed as
\begin{align*}
& \sum_{\bar d \in \cD} \EE[ Y_{t+1} \left((\bar A_{t-1}, a), \bar D = (d_1, d_2) \right)] P(D_1 = d_1, D_2 = d_2) \\
=& \sum_{\bar d \in \{-1, 1\}^2} \EE[ Y_{t+1} \left((\bar A_{t-1}, a), \bar D = (d_1, d_2) \right)] \cdot \frac{1}{2} \cdot \frac{1}{2} \\
=& \frac{1}{4} \left\{ 
\left[ (a - p_{t}(1 \mid \bar Z = (1, 1), R = 0)) (\beta_0^* + \beta_1^* + \beta_2^* + \beta_3^*) + \gamma_0^* + \gamma_1^* + \gamma_2^* + \gamma_3^* \right] P(R = 0 \mid Z_1 = 1) \right. \\
&\ + \left[ (a - p_{t}(1 \mid \bar Z = (1, 0), R = 1)) (\beta_0^* + \beta_1^*) +
\gamma_0^* + \gamma_1^* \right] P(R = 1 \mid Z_1 = 1) \\
& + \left[ (a - p_{t}(1 \mid \bar Z = (1, -1), R = 0)) (\beta_0^* + \beta_1^* - \beta_2^* - \beta_3^*) + \gamma_0^* + \gamma_1^* - \gamma_2^* - \gamma_3^* \right] P(R = 0 \mid Z_1 = 1) \\
&\ + \left[ (a - p_{t}(1 \mid \bar Z = (1, 0), R = 1)) (\beta_0^* + \beta_1^*) +
\gamma_0^* + \gamma_1^* \right] P(R = 1 \mid Z_1 = 1) \\
& + \left[ (a - p_{t}(1 \mid \bar Z = (-1, 1), R = 0)) (\beta_0^* - \beta_1^* + \beta_2^* - \beta_3^*) + \gamma_0^* - \gamma_1^* + \gamma_2^* - \gamma_3^* \right] P(R = 0 \mid Z_1 = -1) \\
&\ + \left[ (a - p_{t}(1 \mid \bar Z = (-1, 0), R = 1)) (\beta_0^* - \beta_1^*) +
\gamma_0^* - \gamma_1^* \right] P(R = 1 \mid Z_1 = -1) \\
& + \left[ (a - p_{t}(1 \mid \bar Z = (-1, -1), R = 0)) (\beta_0^* - \beta_1^* - \beta_2^* + \beta_3^*) + \gamma_0^* - \gamma_1^* - \gamma_2^* + \gamma_3^* \right] P(R = 0 \mid Z_1 = -1) \\
&\ + \left. \left[ (a - p_{t}(1 \mid \bar Z = (-1, 0), R = 1)) (\beta_0^* - \beta_1^*) +
\gamma_0^* - \gamma_1^* \right] P(R = 1 \mid Z_1 = -1) \right\} \\
=& \frac{1}{4} \left\{ 
(a - p_{t}(1 \mid \bar Z = (1, 1), R = 0)) (\beta_0^* + \beta_1^* + \beta_2^* + \beta_3^*) P(R = 0 \mid Z_1 = 1) \right. \\
&\ \ \ + (a - p_{t}(1 \mid \bar Z = (1, -1), R = 0)) (\beta_0^* + \beta_1^* - \beta_2^* - \beta_3^*) P(R = 0 \mid Z_1 = 1) \\
&\ \ \ + (a - p_{t}(1 \mid \bar Z = (-1, 1), R = 0)) (\beta_0^* - \beta_1^* + \beta_2^* - \beta_3^*) P(R = 0 \mid Z_1 = -1) \\
&\ \ \ + (a - p_{t}(1 \mid \bar Z = (-1, -1), R = 0)) (\beta_0^* - \beta_1^* - \beta_2^* + \beta_3^*) P(R = 0 \mid Z_1 = -1) \\
&\ \ \ + 2 (a - p_{t}(1 \mid \bar Z = (1, 0), R = 1)) (\beta_0^* + \beta_1^*) P(R = 1 \mid Z_1 = 1) \\
&\ \ \ + 2 (a - p_{t}(1 \mid \bar Z = (-1, 0), R = 1) (\beta_0^* - \beta_1^*) P(R = 1 \mid Z_1 = -1) \\
&\ \ \ + \left. 4\gamma_0^* \right\}.
\end{align*}
As a result, the marginal effect of the MRT treatment on the proximal outcome averaging over DTRs is
$\sum_{\bar d \in \cD} \EE[ Y_{t+1} \left((\bar A_{t-1}, 1), \bar D = (d_1, d_2) \right) - Y_{t+1} \left((\bar A_{t-1}, 0), \bar D = (d_1, d_2) \right)] P(D_1 = d_1, D_2 = d_2) = \beta_0^*$.

\subsubsection{Marginal Mean at A Fixed DTR Averaging over $A_t$}
\paragraph*{Stage 1}
When $t < t^*$, the marginal mean of the proximal outcome at a fixed DTR averaging over MRT treatments is
\begin{align*}
& \EE \left[ Y_{t+1} \left( \bar A_{t}, D_1 = d_1 \right) \right] \\
=& \sum_{z_1 \in \{-1, 1\}} \sum_{a=0}^1 \EE \left[  I\{z_1 = d_1 (X_0)\} Y_{t+1} \left((\bar A_{t-1}, a), Z_1 = z_1 \right) \right] P(A_t(z_1) = a) \nonumber \\
=& \sum_{z_1 \in \{-1, 1\}} \sum_{a=0}^1 \EE \left[ I\{z_1 = d_1 (X_0)\} \EE \left[ Y_{t+1} \mid A_t = a, Z_1 = z_1 \mid H_t^M \right] \right] P(A_t = a \mid Z_1 = z_1) \nonumber \\
=& \sum_{z_1 \in \{-1, 1\}}  I\{z_1 = d_1 (X_0)\} \left\{ \left[ (1 - p_{t}(1 \mid z_1)) (\beta_0^* + \beta_1^* z_{1}) + \gamma_0^* + \gamma_1^* z_{1} \right] p_{t}(1 \mid z_1) \right. \\
&\ \ + \left. \left[ - p_{t}(1 \mid z_1) (\beta_0^* + \beta_1^* z_{1}) + \gamma_0^* + \gamma_1^* z_{1} \right] (1 - p_{t}(1 \mid z_1)) \right\} \\
=& \sum_{z_1 \in \{-1, 1\}} I\{z_1 = d_1 (X_0)\} \left( \gamma_0^* + \gamma_1^* z_{1} \right).
\end{align*}

\paragraph*{Stage 2}
When $t > t^*$, the marginal mean of the proximal outcome at a fixed DTR averaging over MRT treatments is
\begin{align*}
& \EE \left[ Y_{t+1} \left( \bar Z = (z_1, z_2), \bar A_{t} \right) \right] \\
=& \sum_{z \in \cZ} \sum_{a=0}^1 \sum_{r=0}^1 
\EE \left[ I\{z_1 = d_1 (X_0)\} I\{z_2 = d_2 (H_{t^*}(z_1), R(z_1) = r)\}
Y \left((\bar A_{t-1}, a), \bar Z = (z_1, z_2) \right) \right] \\
&\ \ \ \ \ \ \cdot P(A_t(z_1, r, z_2) = a) P(R(z_1) = r) \nonumber \\
=& \sum_{z \in \cZ} \sum_{a=0}^1 \sum_{r=0}^1 
\EE \left[ I\{z_1 = d_1 (X_0)\} I\{z_2 = d_2 (Z_1 = z_1, R = r)\} \right.\\
&\ \ \ \ \ \ \left. \EE \left[ Y \mid A_{t} = a, \bar Z = (z_1, z_2), R = r, \mid H_t^M \right] \right] 
\cdot P(A_t = a \mid z_1, r, z_2) P(R = r \mid Z_1 = z_1) \nonumber \\
=& \left[ (1 - p_{t}(1 \mid \bar Z = (z_1, z_2), R = 0)) (\beta_0^* + \beta_1^* z_{1} + \beta_2^* z_{2} + \beta_3^* z_{1} z_{2}) +
\gamma_0^* + \gamma_1^* z_{1} + \gamma_2^* z_{2} \right. \\
&\ \ \ \ \left. + \gamma_3^* z_{1} z_{2} \right] \cdot p_{t}(1 \mid \bar Z = (z_1, z_2), R = 0) P(R = 0 \mid Z_1 = z_1) \\
&\ \ + \left[ - p_{t}(1 \mid \bar Z = (z_1, z_2), R = 0) (\beta_0^* + \beta_1^* z_{1} + \beta_2^* z_{2} + \beta_3^* z_{1} z_{2}) + \gamma_0^* + \gamma_1^* z_{1} + \gamma_2^* z_{2} + \gamma_3^* z_{1} z_{2} \right] \\
&\ \ \ \ \cdot (1 - p_{t}(1 \mid \bar Z = (z_1, z_2), R = 0)) P(R = 0 \mid Z_1 = z_1) \\
&\ \ + \left[ (1 - p_{t}(1 \mid \bar Z = (z_1, 0), R = 1)) (\beta_0^* + \beta_1^* z_{1}) +
\gamma_0^* + \gamma_1^* z_{1} \right] \cdot p_{t}(1 \mid \bar Z = (z_1, 0), R = 1) \\
&\ \ \ \ \left. + \gamma_3^* z_{1} z_{2} \right] \cdot p_{t}(1 \mid \bar Z = (z_1, z_2), R = 0) P(R = 1 \mid Z_1 = z_1) \\
&\ \ + \left[ - p_{t}(1 \mid \bar Z = (z_1, 0), R = 1) (\beta_0^* + \beta_1^* z_{1}) +
\gamma_0^* + \gamma_1^* z_{1} \right] \cdot (1 - p_{t}(1 \mid \bar Z = (z_1, 0), R = 1)) \\
&\ \ \ \ \cdot P(R = 1 \mid Z_1 = z_1)\\
=& \left[ \gamma_0^* + \gamma_1^* z_{1} + \gamma_2^* z_{2}+ \gamma_3^* z_{1} z_{2} \right] P(R = 0 \mid Z_1 = z_1) + \left[ \gamma_0^* + \gamma_1^* z_{1} \right] P(R = 1 \mid Z_1 = z_1).
\end{align*}

\newpage
\subsection{Additional Simulation Results} \label{sec:sim:results:additional}

Tables \ref{tab:sim:scenario1:N400} and \ref{tab:sim:scenario2:N400} display results from 500 replications of simulation scenarios I and II, respectively, under sample size $N = 400$.

\begin{table}[H]
\caption{Marginal effect estimation comparisons among three methods in simulation scenario I, where the MRT randomization probability. Sample size $N = 400$.\label{tab:sim:scenario1:N400}}

\begin{subtable}{\linewidth}\centering
    \footnotesize
    \subcaption{\footnotesize{Comparison of effects between $A_t = 1$ and 0, for a fixed DTR or averaging over DTRs.}}\label{tab:sim:results:scenario1:N400:contrastsA}
\begin{tabular}{clllllllll}
\toprule
\multicolumn{1}{c}{ } & \multicolumn{1}{c}{ } & \multicolumn{1}{c}{ } & \multicolumn{1}{c}{ } & \multicolumn{3}{c}{Hybrid} & \multicolumn{3}{c}{WCLS} \\
\cmidrule(l{3pt}r{3pt}){5-7} \cmidrule(l{3pt}r{3pt}){8-10}
 & Stage & Condition & True & Bias & SE & CP & Bias & SE & CP\\
\midrule
1) & 1 & Fix $d_1 = 1$ & 0.1 & 0 & 0.03 & 0.98 & -0.03 & 0.03 & 1\\
2) & 1 & Fix $d_1 = -1$ & 0.7 & 0 & 0.02 & 0.98 & 0.06 & 0.03 & 0.3\\
3) & 2 & Fix $\widebar{d} = (1, 1)$ & 0.14 & 0 & 0.03 & 0.98 & -0.01 & 0.03 & 0.98\\
4) & 2 & Fix $\widebar{d} = (1, -1)$ & 0.06 & 0 & 0.03 & 0.98 & -0.04 & 0.06 & 0.99\\
5) & 2 & Fix $\widebar{d} = (-1, 1)$ & 0.86 & 0 & 0.03 & 0.98 & 0.09 & 0.05 & 0.53\\
6) & 2 & Fix $\widebar{d} = (-1, -1)$ & 0.54 & 0 & 0.03 & 0.97 & 0.04 & 0.03 & 0.81\\
\midrule
7) & 1 & Averaging DTR & 0.4 & 0 & 0.02 & 0.98 & 0.02 & 0.02 & 0.85\\
8) & 2 & Averaging DTR & 0.4 & 0 & 0.02 & 0.98 & 0.02 & 0.02 & 0.85\\
\bottomrule
\end{tabular}
\end{subtable}%
\hspace{0.05\textwidth} 

\begin{subtable}{\linewidth}\centering
    \footnotesize
    \subcaption{\footnotesize{Comparison of effects on the proximal outcome between DTRs averaging over MRT treatments.}}\label{tab:sim:results:scenario1:N400:contrastsZAverageA}
\begin{tabular}{clllllllllll}
\toprule
\multicolumn{1}{c}{ } & \multicolumn{1}{c}{ } & \multicolumn{1}{c}{ } & \multicolumn{1}{c}{ } & \multicolumn{3}{c}{Hybrid} & \multicolumn{3}{c}{WR} \\
\cmidrule(l{3pt}r{3pt}){5-7} \cmidrule(l{3pt}r{3pt}){8-10}
 & Stage & Contrast & True & Bias & SE & CP & Bias & SE & CP & mRE & sdRE\\
\midrule
1) & 1 & $d_1 = 1$ vs -1 & 0.4 & 0 & 0.03 & 0.98 & 0 & 0.03 & 0.99 & 1.20 & 0.06\\
\midrule
2) & 2 & $\widebar{d} = (1, 1)$ vs (1, -1) & -0.16 & 0 & 0.04 & 0.96 & 0 & 0.04 & 0.96 & 1.03 & 0.11\\
3) & 2 & $\widebar{d} = (1, 1)$ vs (-1, 1) & 0.32 & 0 & 0.04 & 0.97 & 0 & 0.04 & 0.98 & 1.05 & 0.06\\
4) & 2 & $\widebar{d} = (1, 1)$ vs (-1, -1) & 0.32 & 0 & 0.04 & 0.98 & 0 & 0.04 & 0.99 & 1.09 & 0.07\\
5) & 2 & $\widebar{d} = (1, -1)$ vs (-1, 1) & 0.48 & 0 & 0.04 & 0.97 & 0 & 0.04 & 0.98 & 1.18 & 0.08\\
6) & 2 & $\widebar{d} = (1, -1)$ vs (-1, -1) & 0.48 & 0 & 0.03 & 0.97 & 0 & 0.04 & 0.98 & 1.24 & 0.08\\
7) & 2 & $\widebar{d} = (-1, 1)$ vs (-1, -1) & 0 & 0 & 0.03 & 0.96 & 0 & 0.03 & 0.96 & 1.05 & 0.06\\
\bottomrule
\end{tabular}
\end{subtable}%
\hspace{0.05\textwidth} 

\begin{subtable}{\linewidth}\centering
    \footnotesize
    \subcaption{\footnotesize{Comparison of effects on the proximal outcome between DTRs for a fixed fixed MRT treatment. }}\label{tab:sim:results:scenario1:N400:contrastsZ:fixedA}
\begin{tabular}{cllllll}
\toprule
\multicolumn{1}{c}{ } & \multicolumn{1}{c}{ } & \multicolumn{1}{c}{ } & \multicolumn{1}{c}{} & \multicolumn{3}{c}{Hybrid} \\
\cmidrule(l{3pt}r{3pt}){5-7}
 & Stage & Contrast & True & Bias & SE & CP\\
\midrule
1) & 1 & $d_1 = 1$ vs -1 & 0.7 & 0 & 0.03 & 0.98\\
2) & 2 & $\widebar{d} = (1, 1)$ vs (1, -1) & -0.2 & 0 & 0.04 & 0.98\\
3) & 2 & $\widebar{d} = (1, 1)$ vs (-1, 1) & 0.68 & 0 & 0.05 & 0.97\\
4) & 2 & $\widebar{d} = (1, 1)$ vs (-1, -1) & 0.52 & 0 & 0.05 & 0.98\\
5) & 2 & $\widebar{d} = (1, -1)$ vs (-1, 1) & 0.88 & 0 & 0.04 & 0.98\\
6) & 2 & $\widebar{d} = (1, -1)$ vs (-1, -1) & 0.72 & 0 & 0.04 & 0.97\\
7) & 2 & $\widebar{d} = (-1, 1)$ vs (-1, -1) & -0.16 & 0 & 0.04 & 0.97\\
\midrule
8) & 1 & $d_1 = 1$ vs -1 & 0.1 & 0 & 0.03 & 0.96\\
9) & 2 & $\widebar{d} = (1, 1)$ vs (1, -1) & -0.12 & 0 & 0.04 & 0.94\\
10) & 2 & $\widebar{d} = (1, 1)$ vs (-1, 1) & -0.04 & 0 & 0.04 & 0.98\\
11) & 2 & $\widebar{d} = (1, 1)$ vs (-1, -1) & 0.12 & 0 & 0.05 & 0.97\\
12) & 2 & $\widebar{d} = (1, -1)$ vs (-1, 1) & 0.08 & 0 & 0.04 & 0.96\\
13) & 2 & $\widebar{d} = (1, -1)$ vs (-1, -1) & 0.24 & 0 & 0.04 & 0.95\\
14) & 2 & $\widebar{d} = (-1, 1)$ vs (-1, -1) & 0.16 & 0 & 0.04 & 0.95\\
\bottomrule
\end{tabular}
\end{subtable}%
\end{table}

\begin{table}[H]
\caption{Marginal effect estimation comparisons among three methods in simulation scenario II, where the MRT randomization probability depends on the DTR assignment. Sample size $N = 400$.\label{tab:sim:scenario2:N400}}

\begin{subtable}{\linewidth}\centering
    \footnotesize
    \subcaption{\footnotesize{Comparison of effects on the proximal outcome between $A_t = 1$ and 0, for a fixed DTR or averaging over DTRs.}}\label{tab:sim:results:scenario2:N400:contrastsA}
\centering
\begin{tabular}{clllllllll}
\toprule
\multicolumn{1}{c}{ } & \multicolumn{1}{c}{ } & \multicolumn{1}{c}{ } & \multicolumn{1}{c}{ } & \multicolumn{3}{c}{Hybrid} & \multicolumn{3}{c}{WCLS} \\
\cmidrule(l{3pt}r{3pt}){5-7} \cmidrule(l{3pt}r{3pt}){8-10}
 & Stage & Condition & True & Bias & SE & CP & Bias & SE & CP\\
\midrule
1) & 1 & Fix $d_1 = 1$ & 0.1 & 0 & 0.03 & 0.94 & -0.02 & 0.03 & 0.9\\
2) & 1 & Fix $d_1 = -1$ & 0.7 & 0 & 0.02 & 0.96 & 0.06 & 0.03 & 0.48\\
3) & 2 & Fix $\widebar{d} = (1, 1)$ & 0.15 & 0 & 0.04 & 0.93 & -0.01 & 0.03 & 0.93\\
4) & 2 & Fix $\widebar{d} = (1, -1)$ & 0.05 & 0 & 0.04 & 0.96 & -0.03 & 0.07 & 0.92\\
5) & 2 & Fix $\widebar{d} = (-1, 1)$ & 0.89 & 0 & 0.04 & 0.96 & 0.08 & 0.05 & 0.69\\
6) & 2 & Fix $\widebar{d} = (-1, -1)$ & 0.51 & 0 & 0.03 & 0.96 & 0.03 & 0.03 & 0.83\\
\midrule
7) & 1 & Averaging DTR & 0.4 & 0 & 0.02 & 0.94 & 0.02 & 0.02 & 0.88\\
8) & 2 & Averaging DTR & 0.4 & 0 & 0.02 & 0.94 & 0.02 & 0.02 & 0.88\\
\bottomrule
\end{tabular}
\end{subtable}%
\hspace{0.05\textwidth} 

\begin{subtable}{\linewidth}\centering
    \footnotesize
    \subcaption{\footnotesize{Comparison of effects on the proximal outcome between DTRs averaging over MRT treatments.}}\label{tab:sim:results:scenario2:N400:contrastsZAverageA}
\centering
\begin{tabular}{clllllllllll}
\toprule
\multicolumn{1}{c}{ } & \multicolumn{1}{c}{ } & \multicolumn{1}{c}{ } & \multicolumn{1}{c}{ } & \multicolumn{3}{c}{Hybrid} & \multicolumn{3}{c}{WR} \\
\cmidrule(l{3pt}r{3pt}){5-7} \cmidrule(l{3pt}r{3pt}){8-10}
 & Stage & Contrast & True & Bias & SE & CP & Bias & SE & CP & mRE & sdRE\\
\midrule
1) & 1 & $d_1 = 1$ vs -1 & 0.4 & 0 & 0.03 & 0.94 & 0 & 0.03 & 0.94 & 1 & 0\\
\midrule
2) & 2 & $\widebar{d} = (1, 1)$ vs (1, -1) & -0.2 & 0 & 0.04 & 0.96 & 0 & 0.04 & 0.96 & 1 & 0\\
3) & 2 & $\widebar{d} = (1, 1)$ vs (-1, 1) & 0.3 & 0 & 0.04 & 0.96 & 0 & 0.04 & 0.96 & 1 & 0\\
4) & 2 & $\widebar{d} = (1, 1)$ vs (-1, -1) & 0.3 & 0 & 0.04 & 0.95 & 0 & 0.04 & 0.95 & 1 & 0\\
5) & 2 & $\widebar{d} = (1, -1)$ vs (-1, 1) & 0.5 & 0 & 0.04 & 0.95 & 0 & 0.04 & 0.95 & 1 & 0\\
6) & 2 & $\widebar{d} = (1, -1)$ vs (-1, -1) & 0.5 & 0 & 0.04 & 0.94 & 0 & 0.04 & 0.94 & 1 & 0\\
7) & 2 & $\widebar{d} = (-1, 1)$ vs (-1, -1) & 0 & 0 & 0.04 & 0.96 & 0 & 0.04 & 0.96 & 1 & 0\\
\bottomrule
\end{tabular}
\end{subtable}%
\hspace{0.05\textwidth} 

\begin{subtable}{\linewidth}\centering
    \footnotesize
    \subcaption{ \footnotesize{Comparison of effects on the proximal outcome between DTRs for a fixed MRT treatment.}}\label{tab:sim:results:scenario2:N400:contrastsZ:fixedA}
\begin{tabular}{cllllll}
\toprule
\multicolumn{1}{c}{ } & \multicolumn{1}{c}{ } & \multicolumn{1}{c}{ } & \multicolumn{1}{c}{} & \multicolumn{3}{c}{Hybrid} \\
\cmidrule(l{3pt}r{3pt}){5-7}
 & Stage & Contrast & True & Bias & SE & CP\\
\midrule
1) & 1 & $d_1 = 1$ vs -1 & 0.62 & 0 & 0.05 & 0.93\\
2) & 2 & $\widebar{d} = (1, 1)$ vs (1, -1) & -0.24 & 0 & 0.06 & 0.96\\
3) & 2 & $\widebar{d} = (1, 1)$ vs (-1, 1) & 0.46 & 0 & 0.05 & 0.97\\
4) & 2 & $\widebar{d} = (1, 1)$ vs (-1, -1) & 0.48 & 0 & 0.05 & 0.96\\
5) & 2 & $\widebar{d} = (1, -1)$ vs (-1, 1) & 0.7 & 0 & 0.05 & 0.95\\
6) & 2 & $\widebar{d} = (1, -1)$ vs (-1, -1) & 0.73 & 0 & 0.06 & 0.94\\
7) & 2 & $\widebar{d} = (-1, 1)$ vs (-1, -1) & 0.03 & 0 & 0.04 & 0.96\\
\midrule
8) & 1 & $d_1 = 1$ vs -1 & 0.02 & 0 & 0.05 & 0.91\\
9) & 2 & $\widebar{d} = (1, 1)$ vs (1, -1) & -0.14 & 0 & 0.05 & 0.95\\
10) & 2 & $\widebar{d} = (1, 1)$ vs (-1, 1) & -0.28 & 0 & 0.06 & 0.94\\
11) & 2 & $\widebar{d} = (1, 1)$ vs (-1, -1) & 0.12 & 0 & 0.05 & 0.91\\
12) & 2 & $\widebar{d} = (1, -1)$ vs (-1, 1) & -0.14 & 0 & 0.05 & 0.93\\
13) & 2 & $\widebar{d} = (1, -1)$ vs (-1, -1) & 0.27 & 0 & 0.04 & 0.93\\
14) & 2 & $\widebar{d} = (-1, 1)$ vs (-1, -1) & 0.41 & 0 & 0.05 & 0.95\\
\bottomrule
\end{tabular}
\end{subtable}%
\end{table}

\newpage
\section{M-Bridge Study Data Construction} \label{sec:data:construction}
The study cohort analyzed in Section \ref{sec:data} was part of a larger trial to inform the development of an API for reducing binge drinking among first-year undergraduates in a large Midwestern University for the 2019–2020 academic year ($N = 891$, 62.4\% female, 76.8\% White; see \cite{patrick2021main} and \cite{carpenter2023SelfrelevantAppeals}, for baseline characteristics). Two-thirds ($N = 591$) of the students involved in the larger trial were randomized to a Stage 1 intervention group at the start of the fall semester, whereas the remaining students were randomized to a control group. Among the 591 students, 295 of them were assigned to early Stage 1 intervention and 296 to late Stage 1 intervention, regarding the timing of receiving personalized normative feedback (PNF). These 591 students were subsequently involved in an MRT design, where they were randomized biweekly to receive one of the two types of self-relevant prompts (self-interest or pro-social) that encouraged completion of four biweekly self-monitoring (SM) surveys of alcohol use. Not all participants were randomized to a prompt at every assessment time. As part of the API, whenever a student was classified as a heavy drinker based on the SM survey response, the invitation to self-relevant prompts stopped and the student was sent a link to an indicated Stage 2 intervention. A total of 158 students eventually entered Stage 2 interventions, depending on when they became heavy drinkers.

Consequently, the proximal outcome, maximum number of alcohol drinks consumed in any 24-hour period between two assessment time points, could be unobserved for two reasons: (1) a student received an MRT prompt but did not complete a SM survey, or (2) a student was flagged as a heavy drinker at a previous assessment time and stopped receiving invitations to SM surveys. For the unobserved proximal outcomes of these individuals, we use their responses to the end-of-semester follow-up (FU) survey as proxy outcomes. In December 2019, the FU survey was distributed to all students, regardless of whether they entered Stage 2, to gather additional drinking information. The FU survey collected timeline followback (TLFB) data about the number of drinks a student drank on each typical day of a week (Sunday through Monday) in the past 30 days. We took the maximum of these number of drinks over seven days as the proxy response for an unobserved proximal outcome at the second, third, and fourth assessment time. However, since the first assessment time preceded the FU survey for more than 30 days, the TLFB data might be an inappropriate proxy for unobserved proximal outcome at the first assessment time. For illustrating our proposed method in a typical hybrid design scenario, we excluded students who did not complete the first SM survey ($N = 117$). We further excluded $N = 38$ students whose unobserved proximal outcomes could not be imputed by TLFB data due to missingness. The final cohort consisted of $N = 428$ students. Table \ref{tab:data:samplesizes} shows the number of students randomized at each assessment time during the MRT design.

\begin{table}[H]
\centering
\caption{Sample size at each assessment time in Stage 1 MRT design analyzed in Section \ref{sec:data}} 
\label{tab:data:samplesizes}
\begin{tabularx}{\textwidth}{YYYY}
  \toprule
Assessment time & Number of students randomized & Number of students who completed SM survey & Number of students flagged as heavy drinkers \\ 
  \midrule
1 & 428 & 428 & 80 \\ 
2 & 348 & 317 & 29 \\ 
3 & 319 & 264 & 13 \\ 
4 & 306 & 237 & 10 \\ 
   \bottomrule
\end{tabularx}
\end{table}








\bibliographystyle{imsart-nameyear} 
\bibliography{references}       

\end{document}